\documentclass[iop]{emulateapj}

\newcommand{\gps}{\ensuremath{g_{\rm P1}}}
\newcommand{\rps}{\ensuremath{r_{\rm P1}}}
\newcommand{\ips}{\ensuremath{i_{\rm P1}}}
\newcommand{\zps}{\ensuremath{z_{\rm P1}}}
\newcommand{\yps}{\ensuremath{y_{\rm P1}}}

\def\ra#1#2#3{#1$^{\rm h}$#2$^{\rm m}$#3$^{\rm s}$}
\def\dec#1#2#3{$#1^\circ#2'#3''$}

\usepackage{epsfig,graphicx}
\usepackage{amsmath}
\usepackage{natbib}
\usepackage{comment}
\usepackage{url}

\begin{document}

\title{Zooming In on the Progenitors of Superluminous Supernovae With the HST}
\submitted{ApJ in press}

\def\cfa{1}
\def\ou{2}
\def\stsci{3}
\def\ua{4}
\def\ein{5}
\def\jhu{6}
\def\uia{7}
\def\uip{8}
\def\uh{9}
\def\ucsc{10}
\def\qub{11}

\author{R.~Lunnan\altaffilmark{\cfa},
 R.~Chornock\altaffilmark{\cfa,\ou},
 E.~Berger\altaffilmark{\cfa},
 A.~Rest\altaffilmark{\stsci},
 W.~Fong\altaffilmark{\cfa,\ua,\ein},
 D.~Scolnic\altaffilmark{\jhu},
 D.~O.~Jones\altaffilmark{\jhu},
 A.~M.~Soderberg\altaffilmark{\cfa},
 P.~M.~Challis\altaffilmark{\cfa},
 M.~R.~Drout\altaffilmark{\cfa},
 R.~J.~Foley\altaffilmark{\cfa,\uia,\uip},
 M.~E.~Huber\altaffilmark{\uh},
 R.~P.~Kirshner\altaffilmark{\cfa},
 C.~Leibler\altaffilmark{\cfa,\ucsc},
 G.~H.~Marion\altaffilmark{\cfa},
 M.~McCrum\altaffilmark{\qub},
 D.~Milisavljevic\altaffilmark{\cfa},
 G.~Narayan\altaffilmark{\cfa},
 N.~E.~Sanders\altaffilmark{\cfa},
 S.~J.~Smartt\altaffilmark{\qub},
 K.~W.~Smith\altaffilmark{\qub}, 
 J.~L.~Tonry\altaffilmark{\uh},
 W.~S.~Burgett\altaffilmark{\uh},
 K.~C.~Chambers\altaffilmark{\uh},
 H.~Flewelling\altaffilmark{\uh},
 R.-P.~Kudritzki\altaffilmark{\uh},
 R.~J.~Wainscoat\altaffilmark{\uh}, and
 C.~Waters\altaffilmark{\uh}
}

\altaffiltext{\cfa}{Harvard-Smithsonian Center for Astrophysics, 60 Garden St., Cambridge, MA 02138, USA}
\altaffiltext{\ou}{Astrophysical Institute, Department of Physics and Astronomy, 251B Clippinger Lab, Ohio University, Athens, OH 45701, USA}
\altaffiltext{\stsci}{Space Telescope Science Institute, 3700 San Martin Dr., Baltimore, MD 21218, USA} 
\altaffiltext{\ua}{Steward Observatory, University of Arizona, 933 North Cherry Avenue, Tucson, AZ 85721, USA}
\altaffiltext{\ein}{Einstein Fellow}
\altaffiltext{\jhu}{Department of Physics and Astronomy, Johns Hopkins University, 3400 North Charles Street, Baltimore, MD 21218, USA}
\altaffiltext{\uia}{Astronomy Department, University of Illinois at Urbana-Champaign, 1002 W. Green Street, Urbana, IL 61801, USA}
\altaffiltext{\uip}{Department of Physics, University of Illinois at Urbana-Champaign, 1010 W. Green Street, Urbana, IL 61801, USA}
\altaffiltext{\uh}{Institute for Astronomy, University of Hawaii, 2680 Woodlawn Drive, Honolulu, HI 96822, USA}
\altaffiltext{\ucsc}{Department of Astronomy and Astrophysics, UCSC, 1156 High Street, Santa Cruz, CA 95064, USA}
\altaffiltext{\qub}{Astrophysics Research Centre, School of Mathematics and Physics, Queen's University Belfast, Belfast BT7 1NN, UK}
\email{rlunnan@cfa.harvard.edu}

\begin{abstract}
We present {\it Hubble Space Telescope} (\textit{HST})rest-frame ultraviolet imaging of the host galaxies of 16 hydrogen-poor superluminous supernovae (SLSNe), including 11 events from the Pan-STARRS Medium Deep Survey. Taking advantage of the superb angular resolution of {\it HST}, we characterize the galaxies' morphological properties, sizes and star formation rate (SFR) densities. We determine the supernova (SN) locations within the host galaxies through precise astrometric matching, and measure physical and host-normalized offsets, as well as the SN positions within the cumulative distribution of UV light pixel brightness. We find that the host galaxies of H-poor SLSNe are irregular, compact dwarf galaxies, with a median half-light radius of just 0.9~kpc. The UV-derived SFR densities are high ($\langle \Sigma_{\rm SFR} \rangle \simeq 0.1 {\rm M}_{\odot}{\rm yr}^{-1} {\rm kpc}^{-2}$), suggesting that SLSNe form in overdense environments. Their locations trace the UV light of their host galaxies, with a distribution intermediate between that of long-duration gamma-ray bursts (LGRBs) (which are strongly clustered on the brightest regions of their hosts) and a uniform distribution (characteristic of normal core-collapse SNe), though cannot be statistically distinguished from either with the current sample size. Taken together, this strengthens the picture that SLSN progenitors require different conditions than those of ordinary core-collapse SNe to form, and that they explode in broadly similar galaxies as do LGRBs. If the tendency for SLSNe to be less clustered on the brightest regions than are LGRBs is confirmed by a larger sample, this would indicate a different, potentially lower-mass progenitor for SLSNe than LRGBs.
\end{abstract}


\section{Introduction}
\label{sec:intro}

Superluminous supernovae (SLSNe) are a rare class of supernovae (SNe) discovered in wide-field surveys in the past decade. They are characterized by peak luminosities of 10-100 times those of normal core-collapse and Type Ia SNe. At least two clear subclasses have emerged: SLSNe that show narrow hydrogen lines in their spectra (H-rich SLSNe) are thought to represent the extreme end of the Type IIn SN distribution, and are likely powered by interaction with dense circumstellar medium (CSM) \citep[e.g.][]{ock+07,slf+07, scs+10, cwv+11, ddm+11b, rfg+11}. For the subclass of SLSNe without hydrogen in their spectra (H-poor SLSNe; e.g. \citealt{qaw+07, qkk+11, bdt+09,ccs+11, psb+10, lcd+12, lcb+13, isj+13}), the mechanism powering the extreme luminosities is not known.  As in the case of H-rich SLSNe, interaction has also been proposed as the power source, but would require extreme mass loss and should produce narrow lines in the spectra that are not seen \citep{ci11, gb12, cw12b, mbt+13, bnc+14, nsj+14}. A central engine model, such as energy injection from a newborn magnetar \citep[e.g.][]{woo10,kb10}, has also been proposed. Another possibility, applicable to the slowest-evolving H-poor SLSNe, is pair-instability SNe (PISNe; \citealt{gmo+09,gal12,kwh11,cw12a,kbl+14}), although this interpretation is controversial \citep{ysv+10,mtt+10,dhw+12,dwl+13,nsj+13}.

 One way to shed light on the nature of these extreme explosions is to study their progenitor populations. As all SLSNe discovered to date are far too distant for direct progenitor detections, in practice this means studying their host galaxy environments as a proxy. Early studies \citep{nsg+11, sps+11} suggested that SLSNe were preferentially found in low-luminosity host galaxies. Detailed studies of two individual hosts \citep{csb+13,lcb+13} revealed metal-poor dwarf galaxies with high specific star formation rates. These initial trends were investigated in detail with a much larger sample in \citet{lcb+14}, who studied properties of 31 SLSN host galaxies, and found that compared to core-collapse SNe, the SLSNe are found in lower-luminosity, lower-mass, higher specific star formation rates (sSFRs) and lower metallicity environments. Instead, their properties were found to be broadly consistent with those of long-duration gamma-ray burst (LGRB) host galaxies, though the SLSN host galaxies studied were even lower-luminosity and lower-mass than the LGRB host comparison sample. 

A complementary approach is to analyze the sub-galactic environments of SLSNe, i.e. the locations of SLSNe within their host galaxies. Studies that compare the locations of Type Ic SNe and LGRBs to the sites of star formation, as traced by UV or H$\alpha$ emission, show that the supernova locations are strongly correlated with the brightest star-forming regions \citep{fls+06, slt+10, kkp08, ja06, ahj+12}. This is used as evidence that the progenitors of these explosions are young and massive. Type II SNe overall trace the UV emission, though not as strongly as Type Ib/c SNe, suggesting a longer-lived and less massive progenitor \citep{ahj+12, kkp08}. By contrast, the locations of short-duration gamma-ray bursts are unassociated with UV light and show significant offsets from the host galaxy centers, suggesting they do not arise from young, massive stars, and are consistent instead with the predictions for a compact object merger progenitor \citep{fbf10, fb13}. Thus, studying the locations of transients within their (resolved) host light distributions offers a powerful probe of the progenitor properties.

Here, we present the first analysis of the locations of H-poor SLSNe within their host galaxies. We present resolved {\it Hubble Space Telescope} ({\it HST}) imaging, allowing us to both study the morphologies of SLSN host galaxies, as well as the host-SN offsets and positions of the SN locations in the overall light distribution. We present our targets, {\it HST} observations, data processing techniques and astrometric matching to determine the SLSN locations in Section~\ref{sec:obs}. Section~\ref{sec:stats} describes the comparison samples and statistical techniques, and we present our results in Sections~\ref{sec:galscale} and \ref{sec:results}. The implications of our findings are discussed in Section~\ref{sec:disc} and summarized in Section~\ref{sec:conc}. All calculations in this paper assume a $\Lambda$CDM cosmology with $H_0 = 70$~km~s$^{-1}$~Mpc$^{-1}$, $\Omega_{\rm M} = 0.27$ and $\Omega_{\Lambda} = 0.73$ \citep{ksd+11}.

\section{Observations and Data Analysis}
\label{sec:obs}

\subsection{SLSNe Discovered in the Pan-STARRS Medium Deep Survey (PS1/MDS)}
\label{sec:ps1mds}

The majority of our targets for this study were discovered in the PS1/MDS transient search, which operated from late 2009 to early 2014.
The PS1 telescope on Haleakala is a high-etendue wide-field survey
instrument with a 1.8-m diameter primary mirror and a $3.3^\circ$
diameter field of view imaged by an array of sixty $4800\times 4800$
pixel detectors, with a pixel scale of $0.258''$
\citep{PS1_system,PS1_GPCA}.   \citet{tsl+12} describes the photometric system and broadband filters in detail. 

The PS1/MDS consists of 10 fields (each with a
single PS1 imager footprint) observed in \gps\rps\ips\zps with a typical cadence of 3~d in each filter, to a $5\sigma$ depth of $\sim 23.3$ mag; \yps is used near full moon with a typical depth of $\sim 21.7$ mag.
The standard reduction, astrometric solution, and stacking of the
nightly images are done by the Pan-STARRS1 Image Processing Pipeline (IPP) system \citep{PS1_IPP, PS1_astrometry} on a computer cluster at the Maui High Performance Computer Center. For the transients search, the nightly MDS stacks were transferred
to the Harvard FAS Research Computing cluster, where they were
processed through a frame subtraction analysis using the {\tt photpipe}
pipeline developed for the SuperMACHO and ESSENCE surveys \citep{rsb+05, gsc+07, mpr+07,rsf+14}.

A subset of targets was selected for spectroscopic follow-up, using the Blue Channel spectrograph on the 6.5-m MMT telescope \citep{swf89}, the Gemini Multi-object Spectrograph (GMOS; \citealt{hja+04}) on the 8-m Gemini telescopes, and the Low Dispersion Survey Spectrograph (LDSS3) and Inamori-Magellan Areal Camera and Spectrograph (IMACS; \citealt{dhb+06}) on the 6.5-m Magellan telescopes. Over the 4 yr of the survey, we have discovered and spectroscopically confirmed more than 15 H-poor SLSNe in the PS1/MDS data (\citealt{ccs+11,bcl+12,cbr+13,lcb+13,msr+14,msk+14}; R. Lunnan et al. 2015, in preparation). Due to the modest area and deep detection limit, most of the volume of PS1/MDS is at high redshift. The rare SLSNe in our sample cover $ 0.5 \lesssim z \lesssim 1.6$. 

The spectroscopic follow-up of PS1/MDS was not complete, with follow-up targets selected based on the available light curve and galaxy information. In particular, the SLSNe were generally found by some combination of having long observed rise-times and/or being several magnitudes brighter than any apparent host. \citet{lcb+14} examined in detail to what extent the selection could bias the resulting host galaxy population of the PS1/MDS SLSN sample, with the conclusion that the strong environmental preferences seen are real and not caused by selection effects.

\subsection{HST Observations}
\label{sec:hst}
We obtained {\it HST} observations of the host galaxies of 11 H-poor SLSNe discovered in the PS1/MDS survey through programs GO-13022 and GO-13326 (PIs: Berger and Lunnan, respectively). The initial program targeted five SLSN host galaxies that were undetected in ground-based data, obtaining both rest-frame UV and rest-frame optical imaging, and the follow-up program added rest-frame UV imaging of the remaining SLSN host galaxies in the PS1/MDS sample at the time. In addition the host galaxy of SLSN PS1-10bzj \citep{lcb+13} has archival {\it HST} imaging from the GEMS survey \citep{rbb+04}. Since the remaining galaxy images by necessity were obtained after the SN explosion, the programs were designed to only include targets where the SNe were expected to have faded well below {\it HST} detection threshold by the time of the observations, based on the available PS1/MDS light curves. For this reason, the sample only includes events from the first 2.5 yr of PS1/MDS. All targets are listed in Table~\ref{tab:targets}.

 Each galaxy was imaged with the Advanced Camera for Surveys (ACS/WFC) in a filter corresponding to rest-frame UV emission ($\simeq 3000$\AA; F475W, F606W or F814W were used depending on the redshift). We used a standard four-point dither pattern for optimal pixel subsampling; Table~\ref{tab:hst} lists the details for each observation, including the effective wavelengths for each filter/redshift combination. We processed and combined the calibrated and CTE-corrected individual images using the Astrodrizzle software provided by STScI \citep{ghf+12}, with a final {\tt pixscale} of 0.025\arcsec/pixel (i.e. half the native image scale), and a {\tt pixfrac} value of 0.8.

\begin{figure*}
\begin{center}
\begin{tabular}{cccc}
\includegraphics[width=4.0cm]{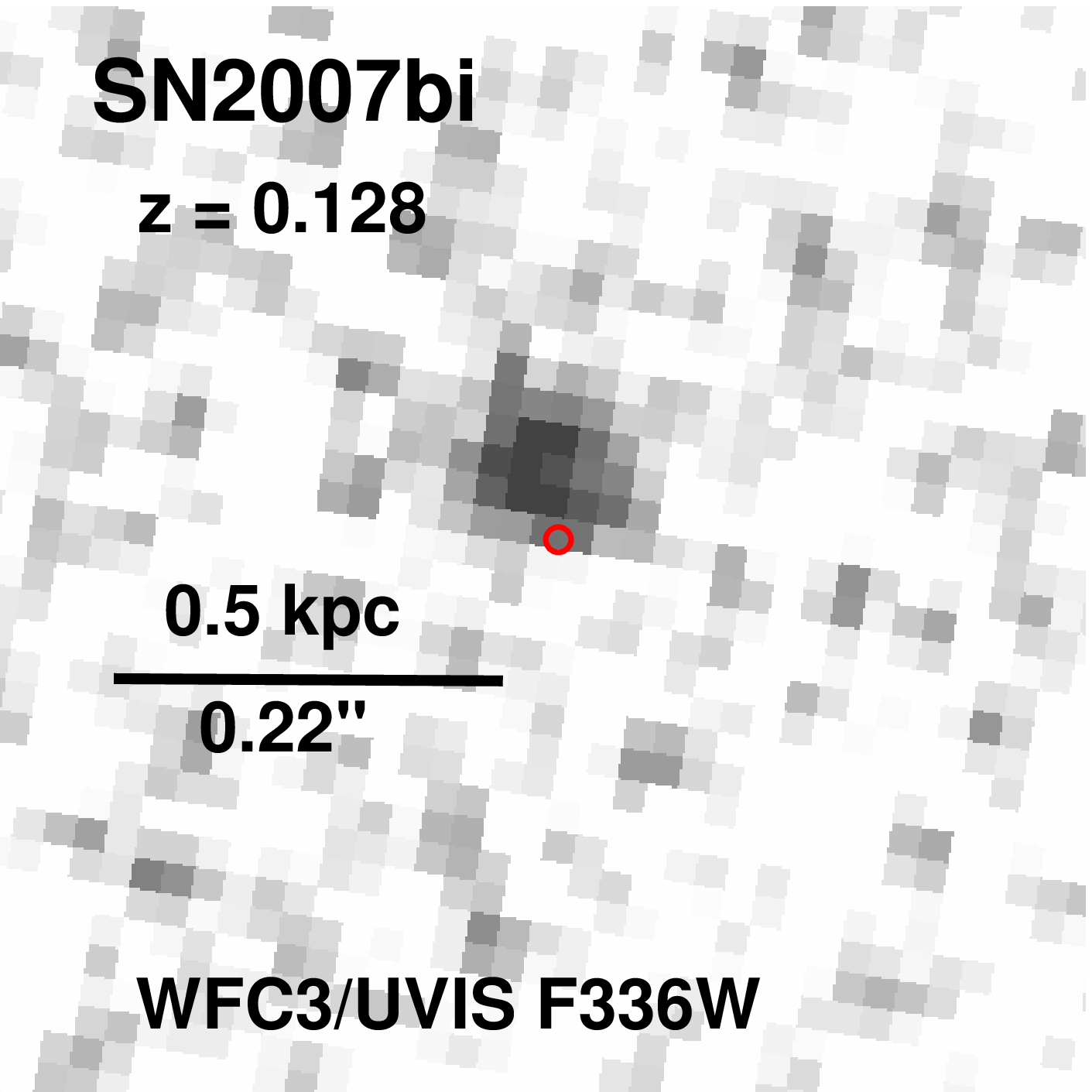} &\includegraphics[width=4.0cm]{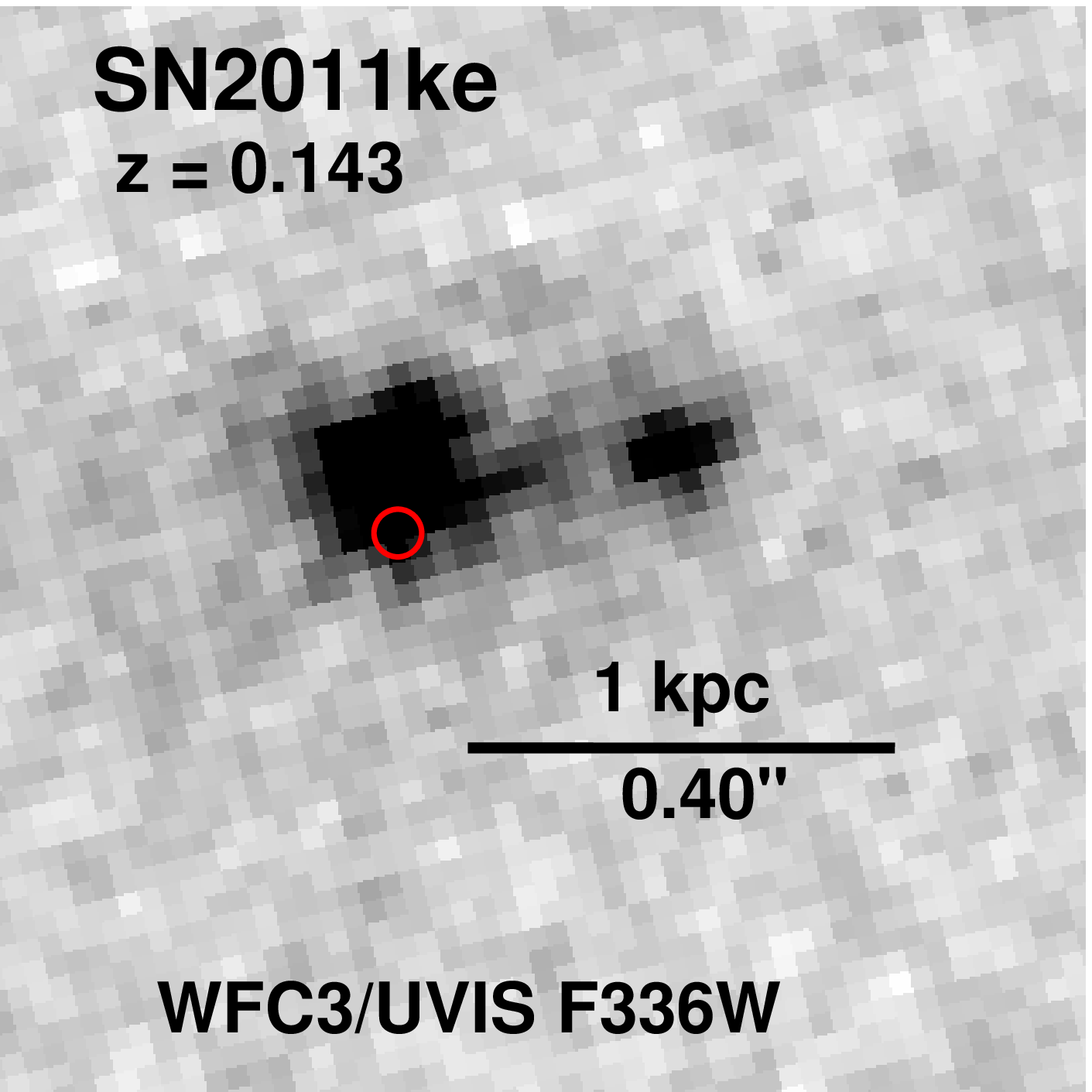} & 
\includegraphics[width=4.0cm]{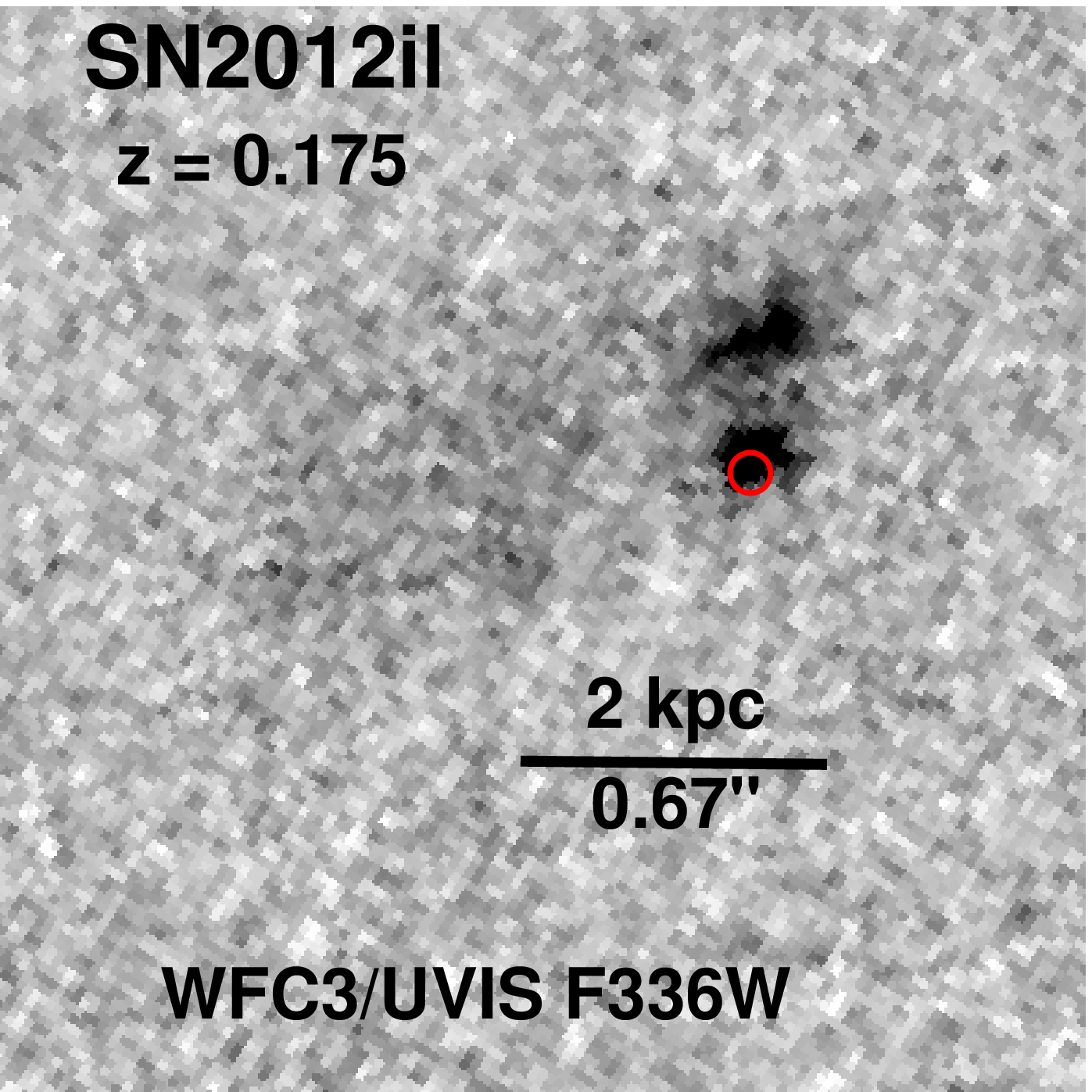} & \includegraphics[width=4.0cm]{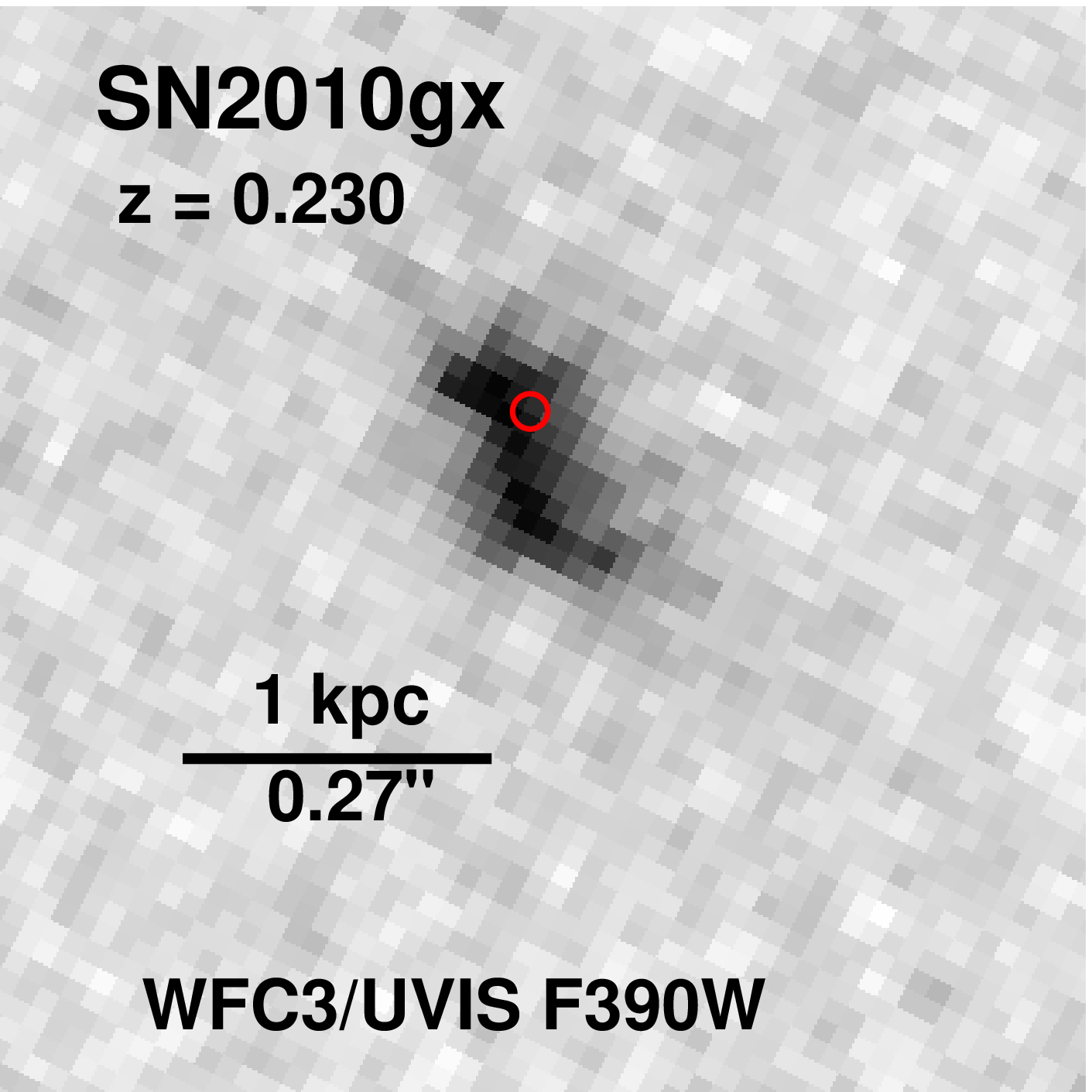} \\
\includegraphics[width=4.0cm]{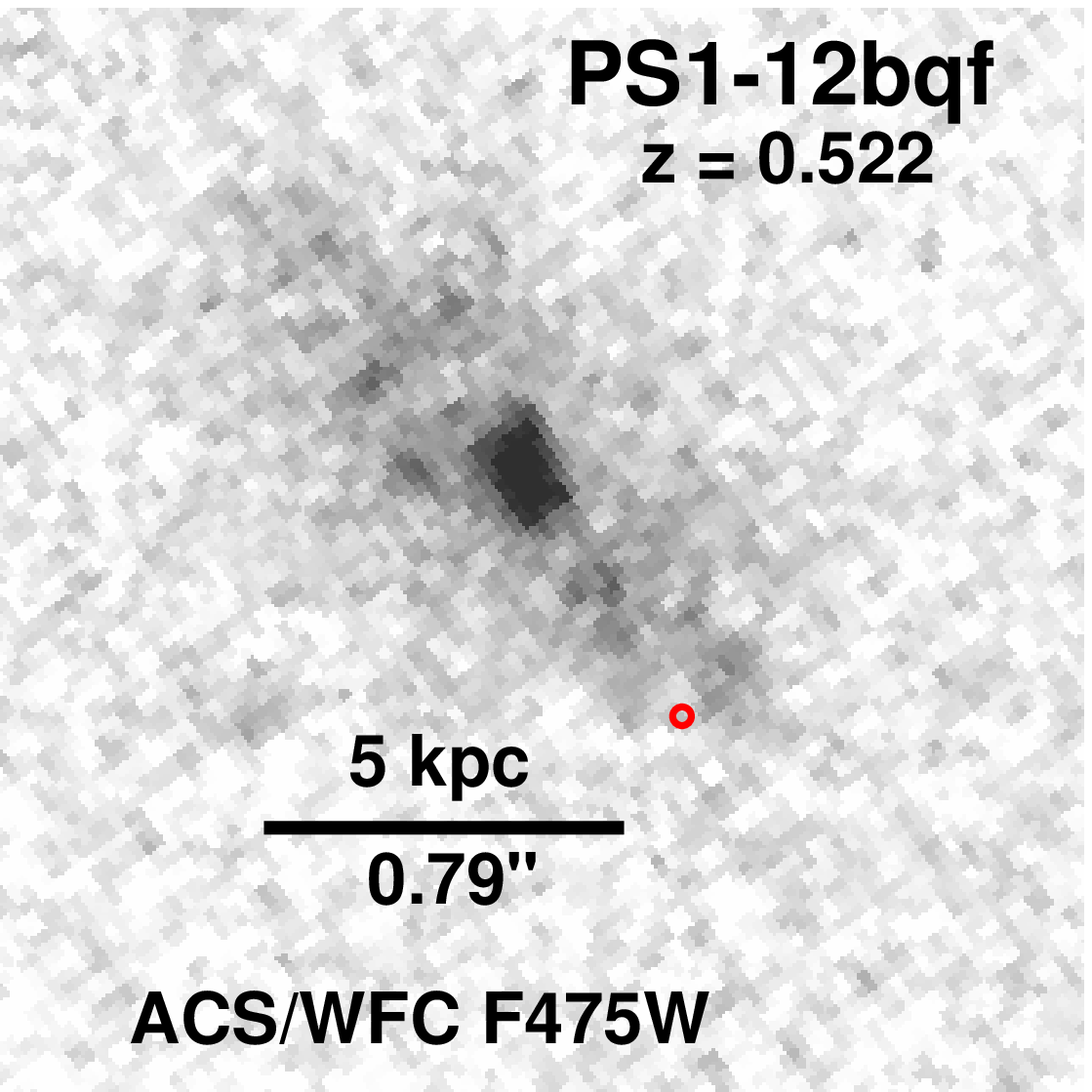} & \includegraphics[width=4.0cm]{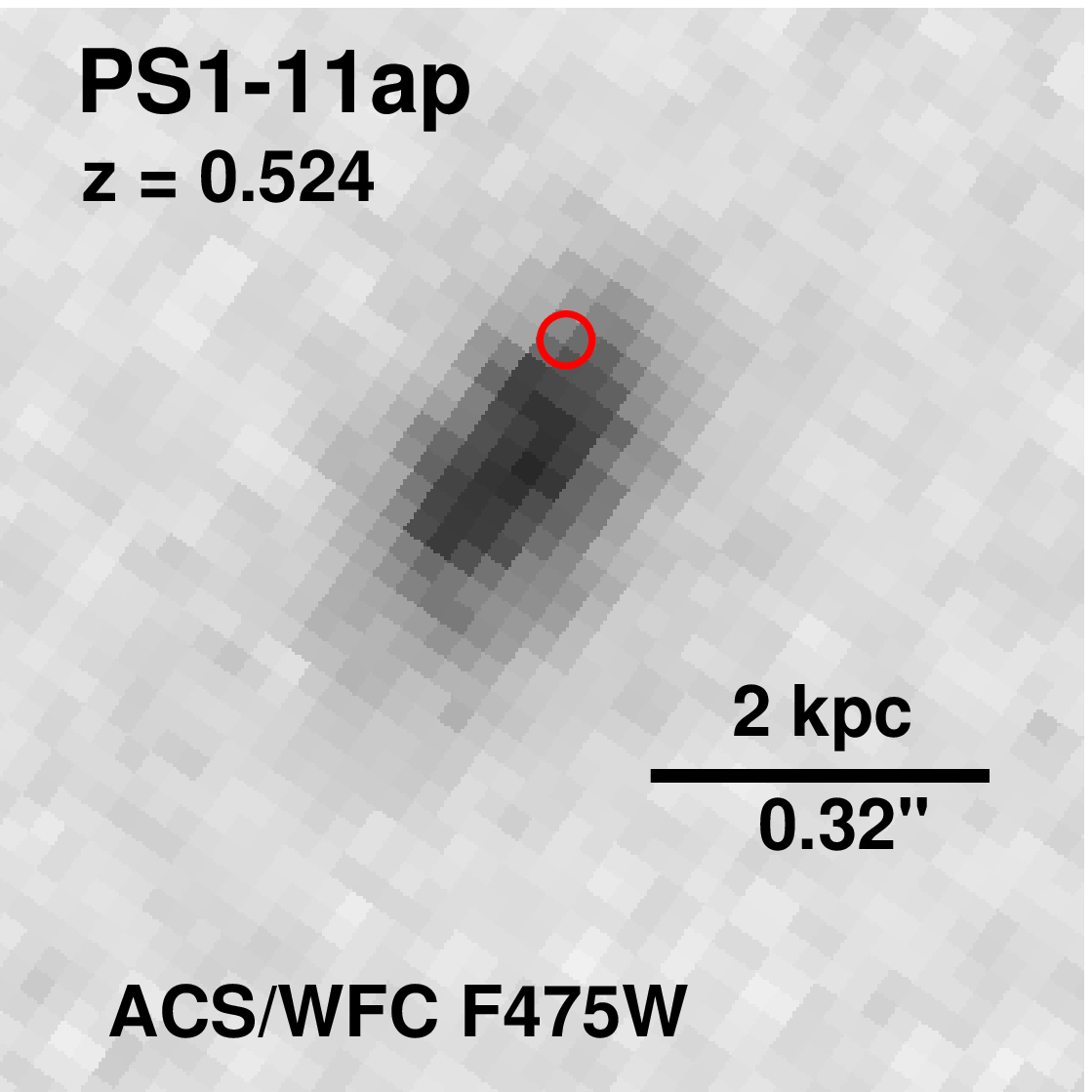} & 
\includegraphics[width=4.0cm]{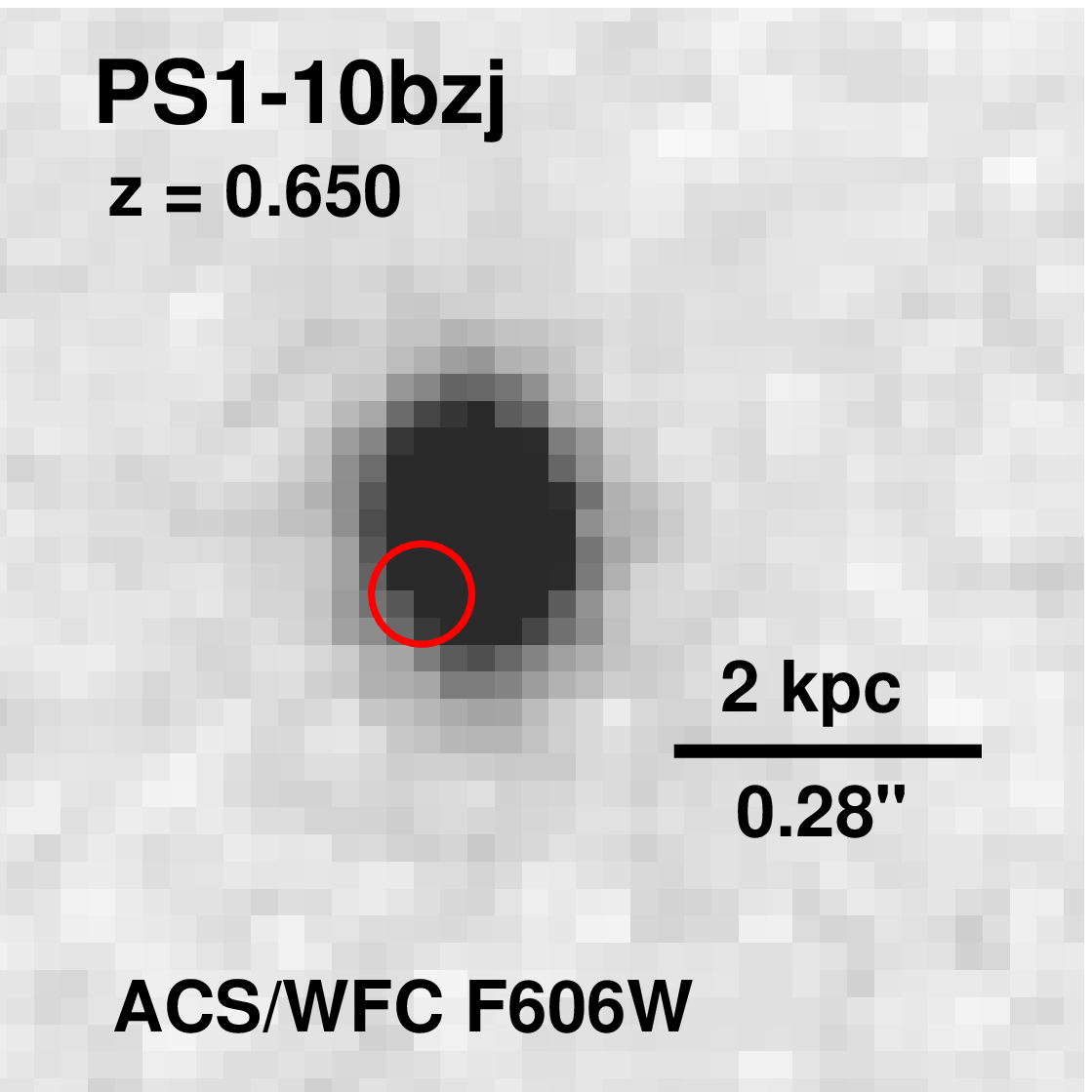} & \includegraphics[width=4.0cm]{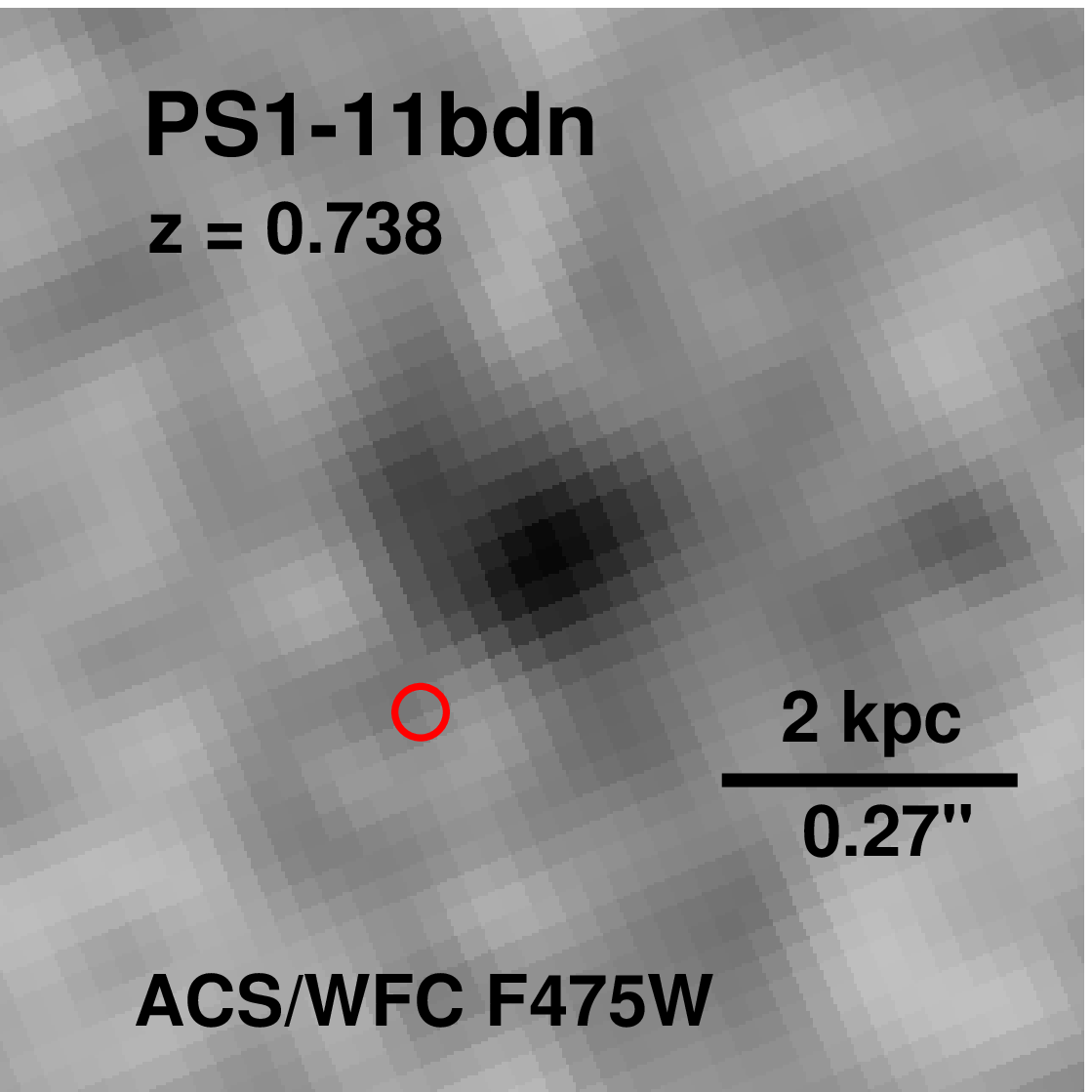} \\
\includegraphics[width=4.0cm]{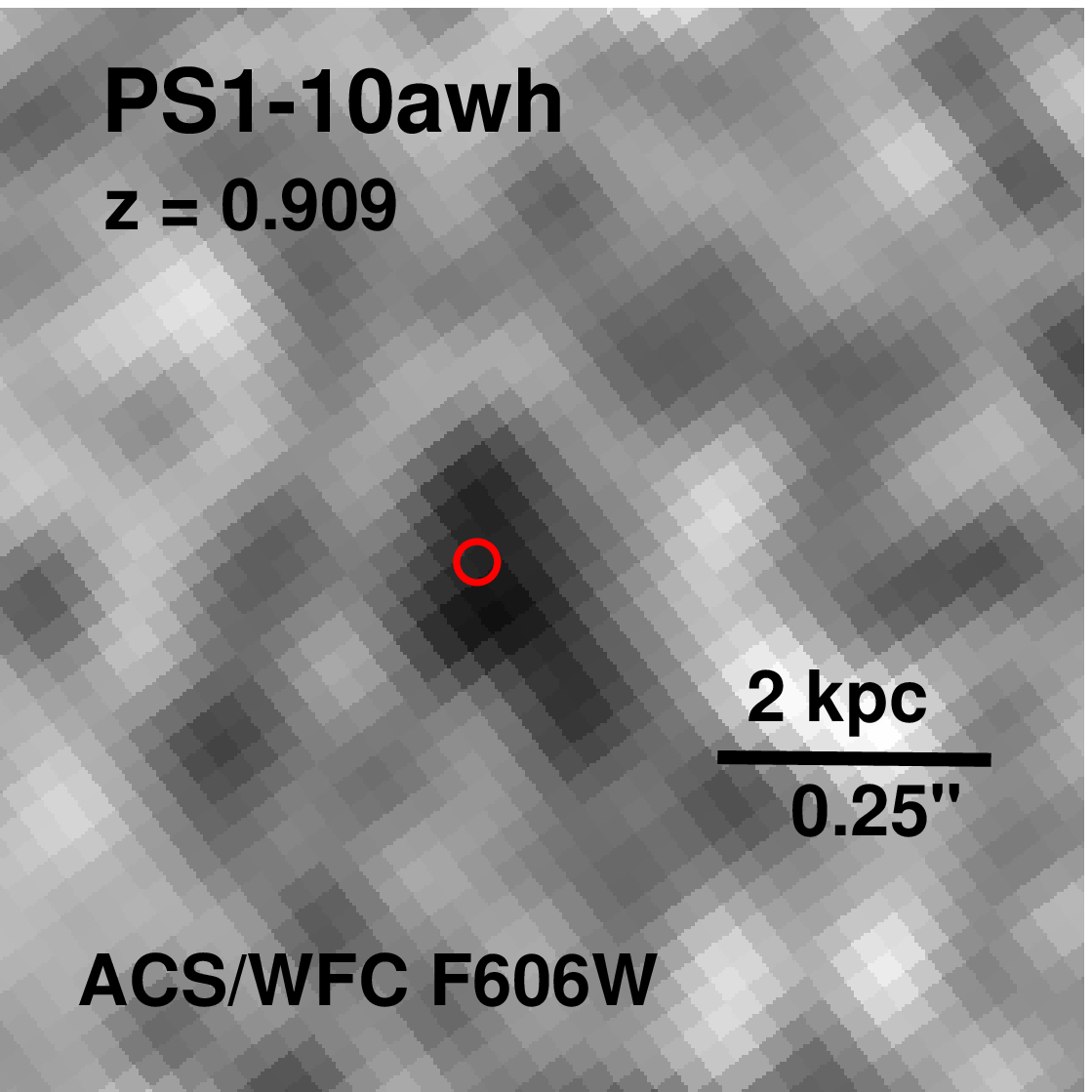}  & \includegraphics[width=4.0cm]{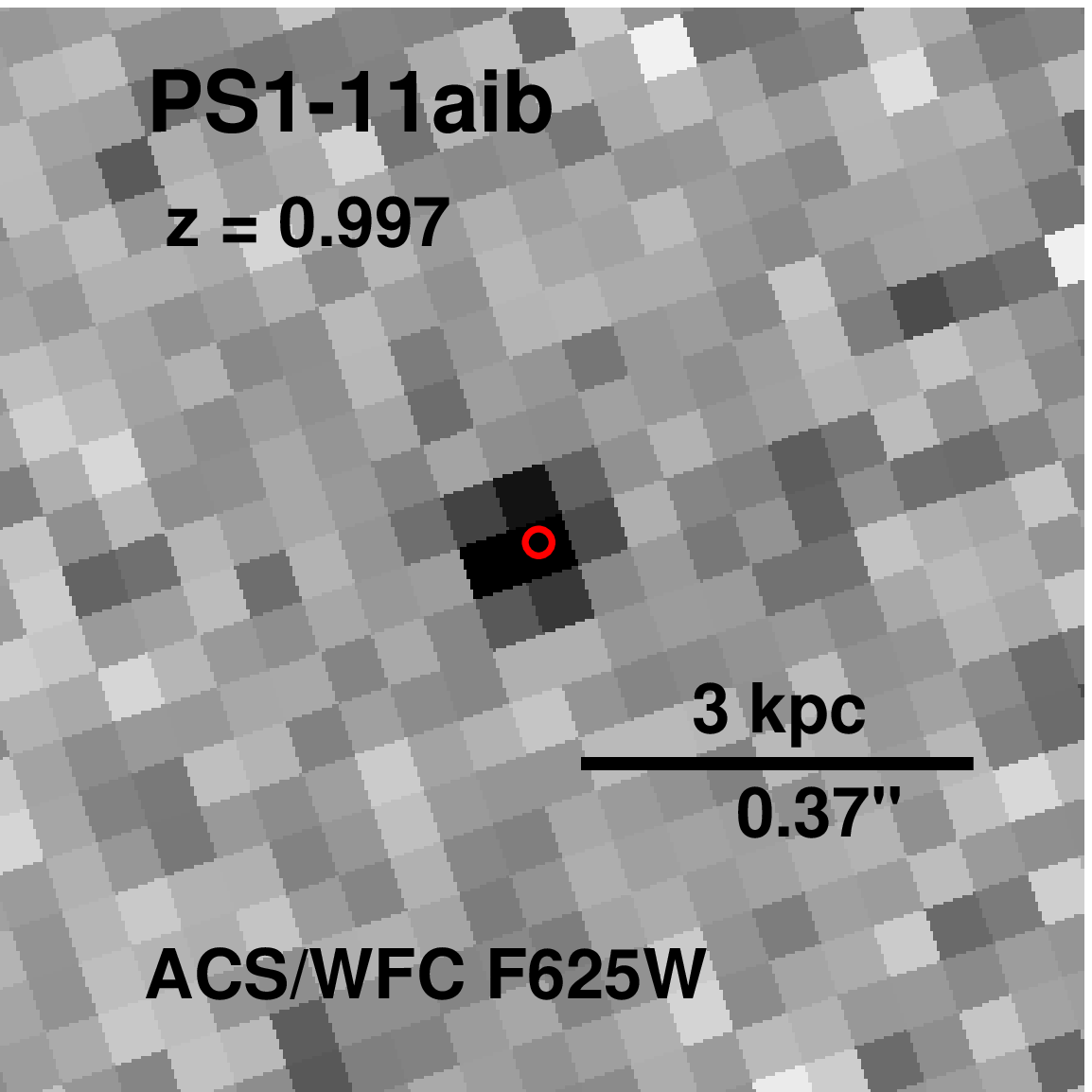} & 
\includegraphics[width=4.0cm]{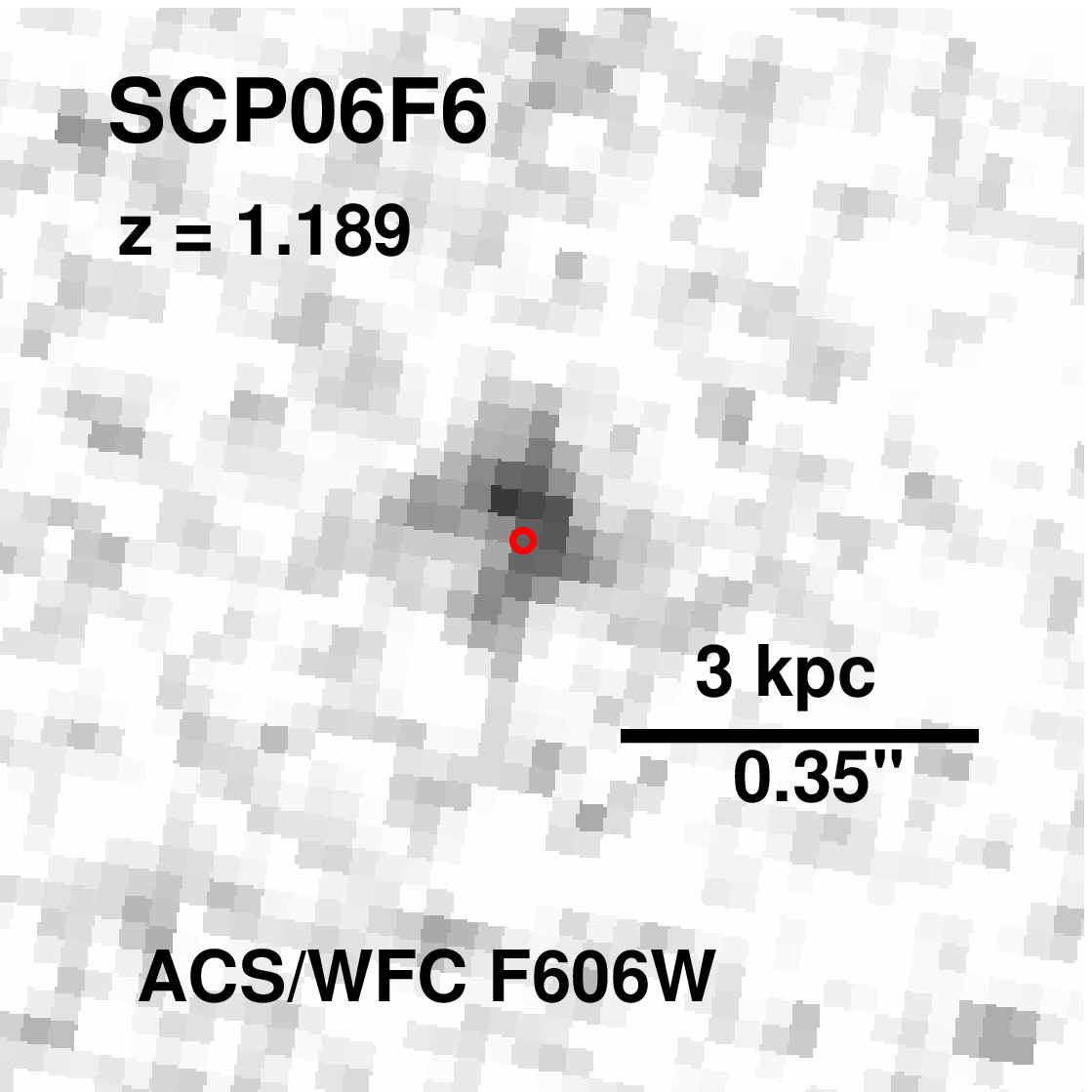}  & \includegraphics[width=4.0cm]{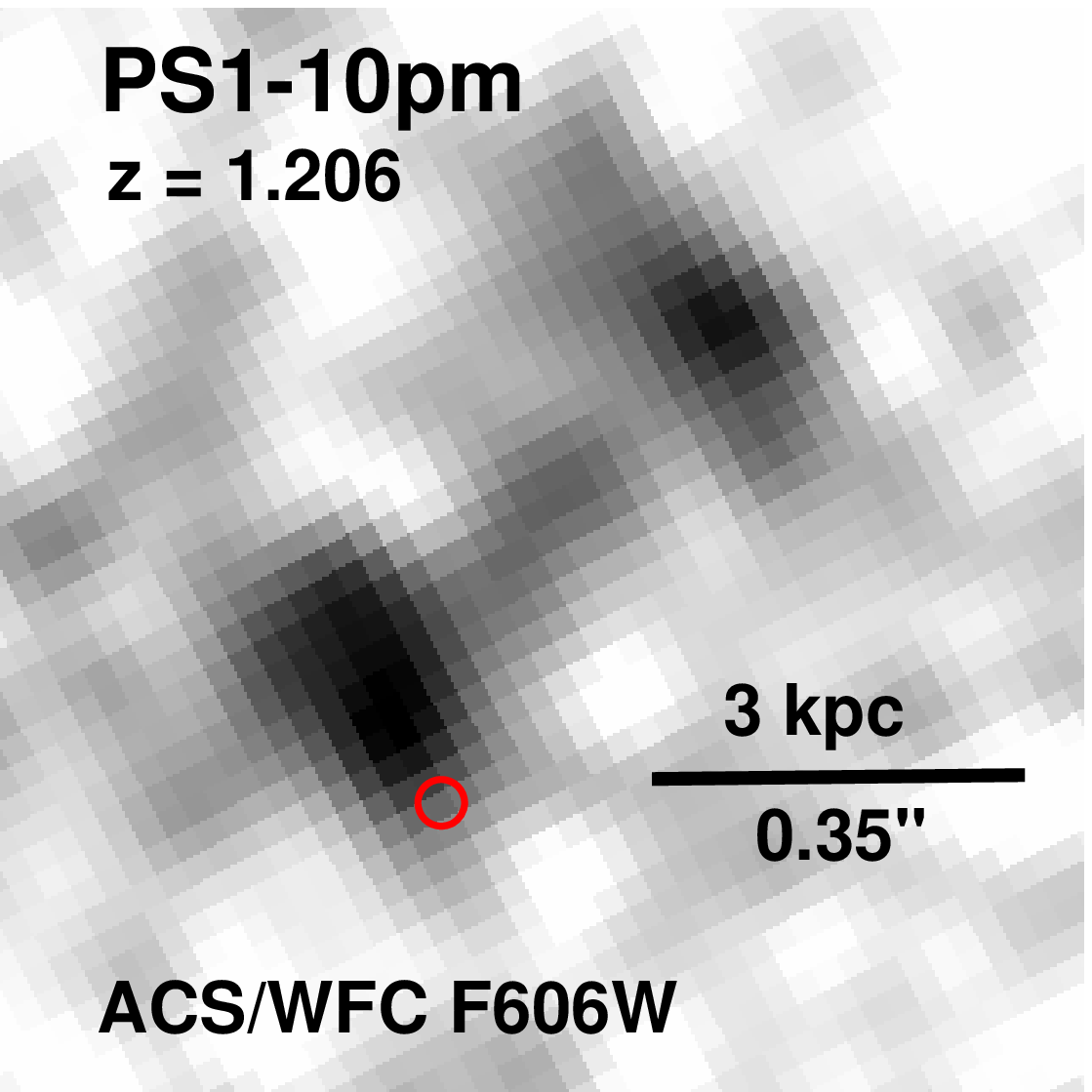} \\ 
\includegraphics[width=4.0cm]{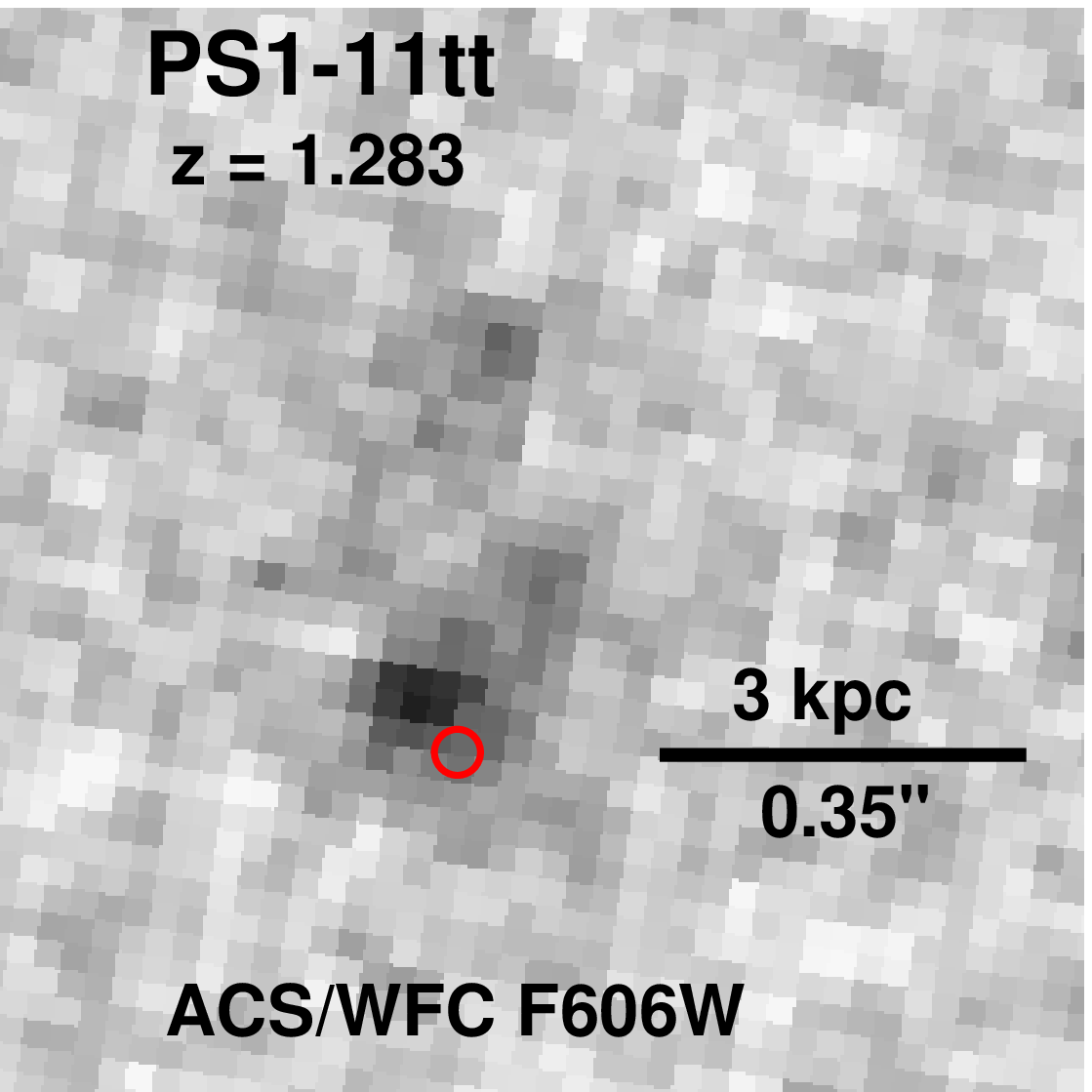} & \includegraphics[width=4.0cm]{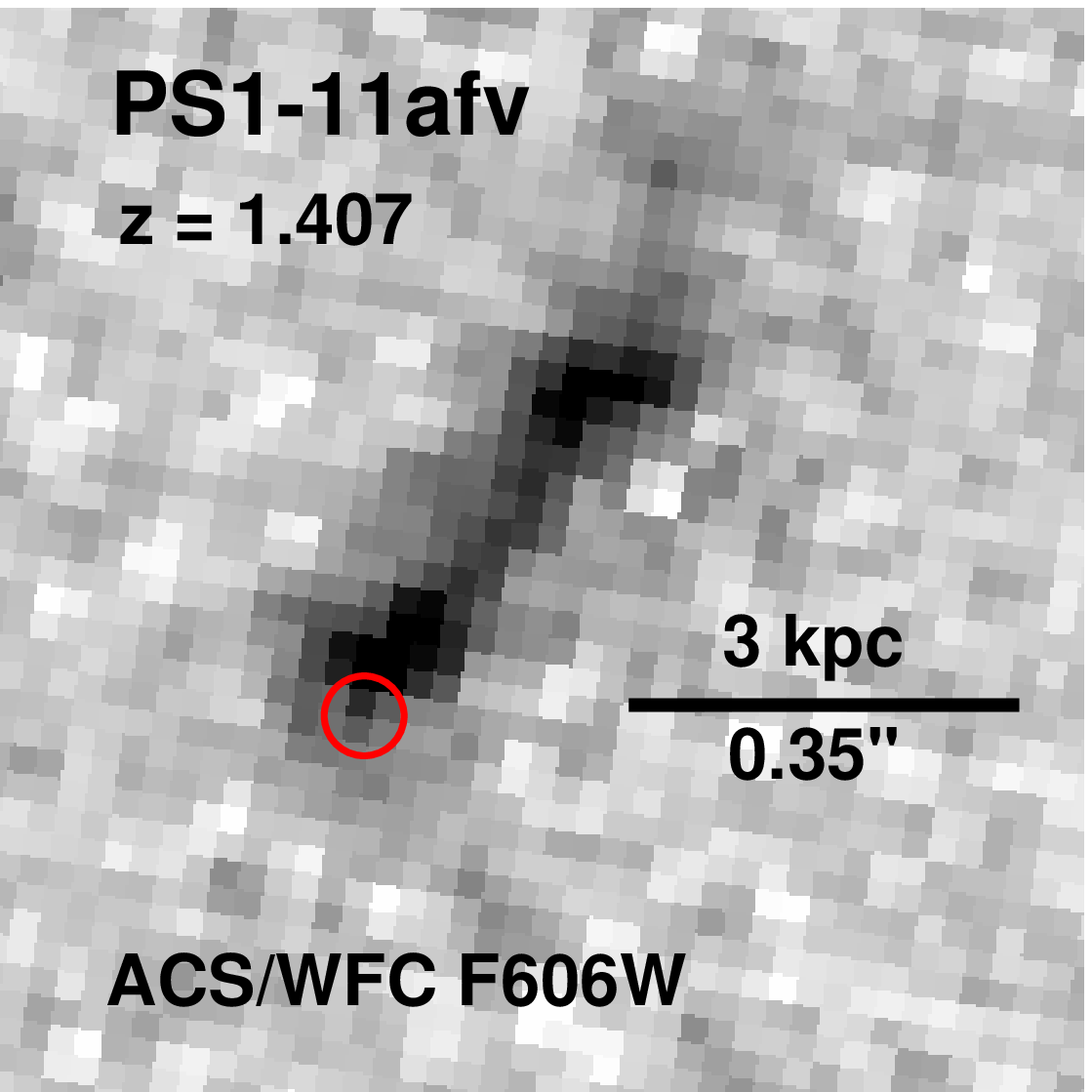}  & 
\includegraphics[width=4.0cm]{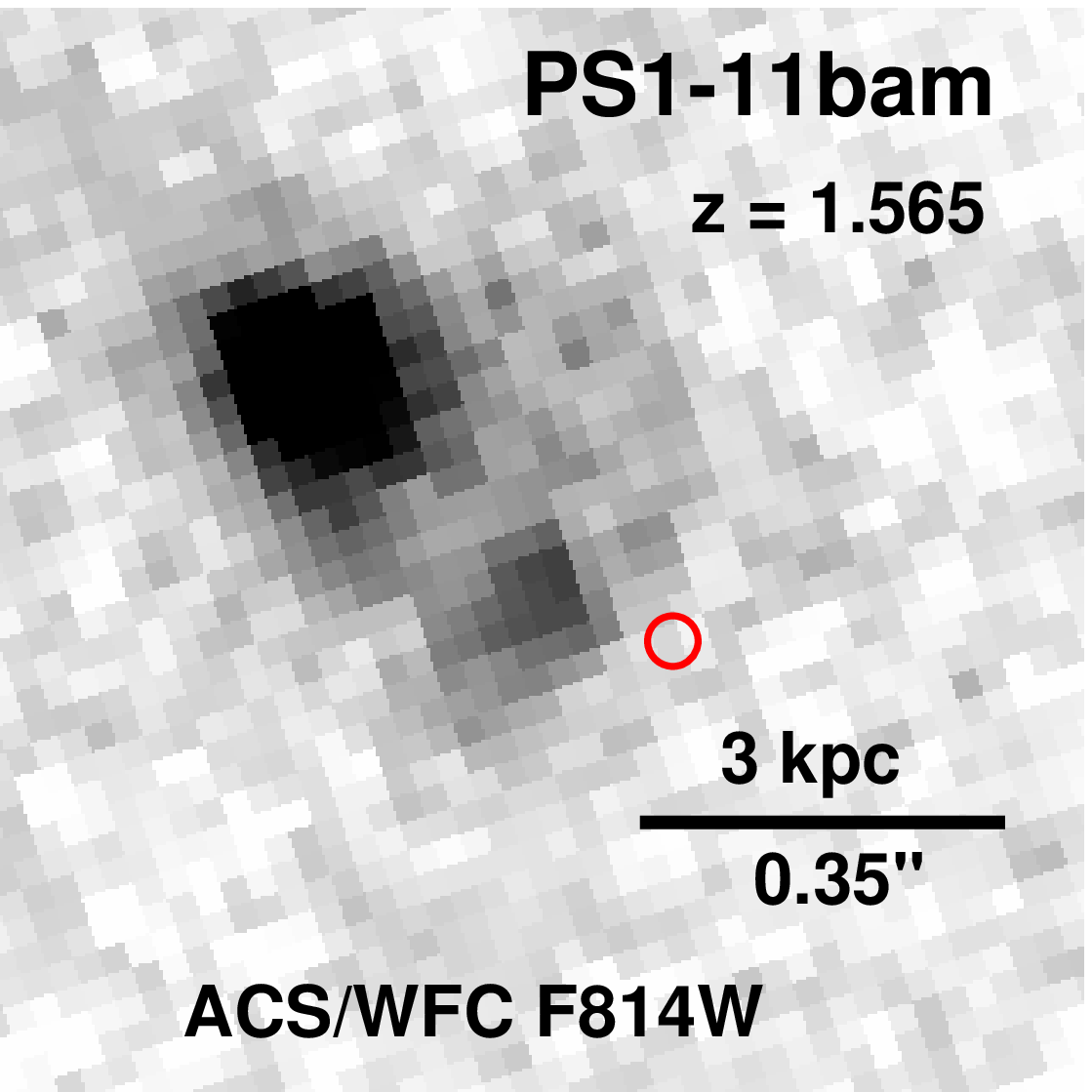} &  \includegraphics[width=4.0cm]{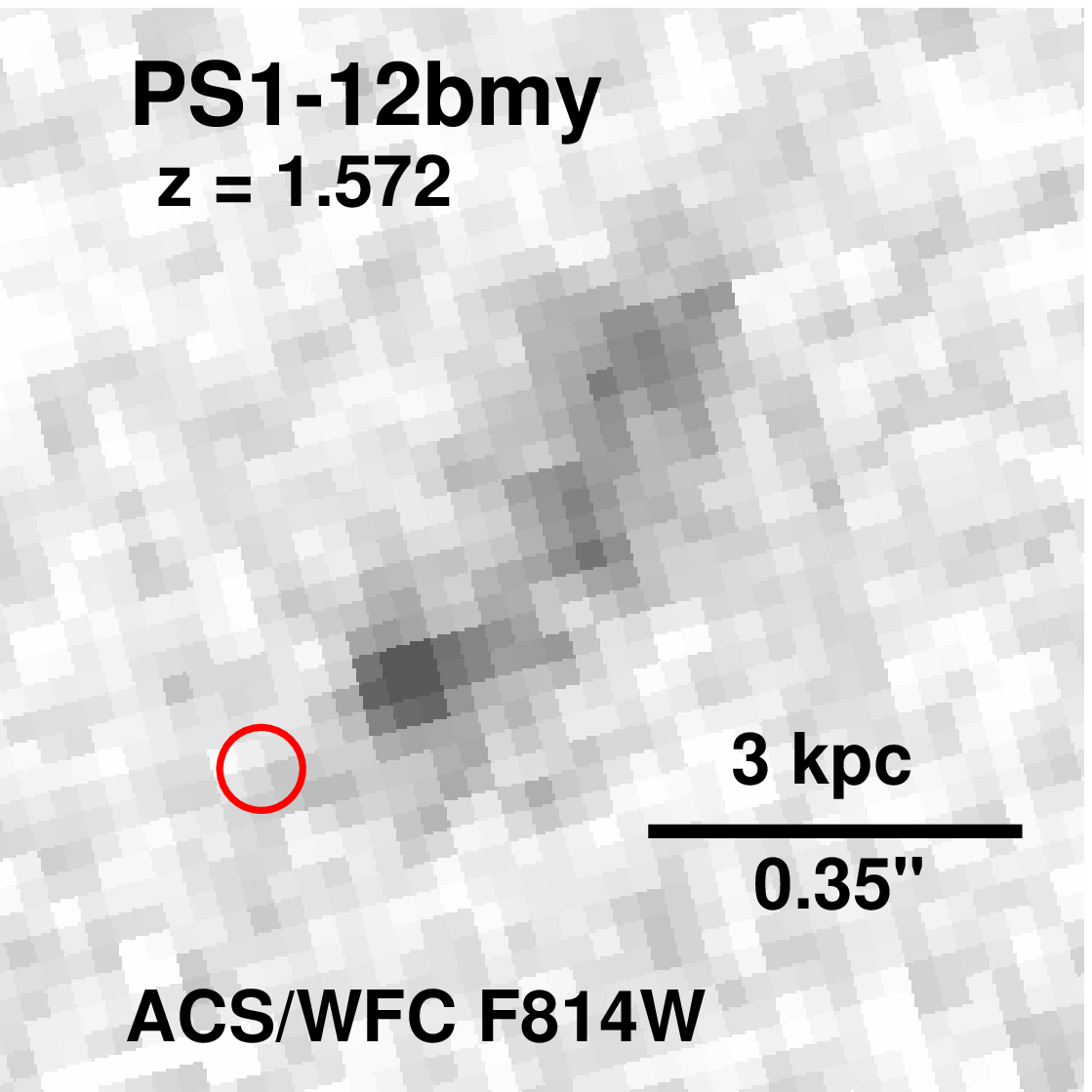} 
\end{tabular}
\caption{{\it HST} rest-frame UV images of 11 host galaxies of PS1/MDS SLSNe, and five host galaxies from the literature. All images are oriented with north up and east pointing left. The horizontal bars show the scale of each image. The red circles correspond to the 1$\sigma$ astrometric uncertainty in the SN position relative to the {\it HST} image, as described in Section~\ref{sec:astrometry}. Some images have been smoothed with a 3 pixel Gaussian filter to make the galaxy more apparent.
 }
\label{fig:hstpix}
\end{center}
\end{figure*}

In addition, the H-poor SLSN PS1-11aib was observed in program GO-12529 (PI: Soderberg), capturing both the late-time light curve of the SN and a final epoch for a host galaxy template. The F625W filter corresponds to rest-frame UV at a redshift of $z=0.997$. In this case, the data are somewhat shallower and we only have two images available per filter, so we do not redrizzle onto a finer grid but keep the original image scale of 0.05\arcsec/pixel. We note that what we identify as the host galaxy of PS1-11aib is unresolved in the template image, and does not appear to be offset from the SN centroid (Figure~\ref{fig:hstpix}), so there is a possibility of confusion with lingering SN emission. The fact that the F625W flux remained constant in the two final epochs (220 to 350 rest-frame days past peak), as well as the flat F625W-F775W color in the final epoch argues that we are indeed detecting the galaxy, however.

The peculiar transient PS1-10afx \citep{cbr+13} was targeted as part of program GO-13326; however the discovery of a second galaxy along the line of sight combined with the observed SN properties make it likely this object was in fact a lensed SN rather than a SLSN \citep{qwo+13,qom+14}. Due to the uncertainties in the nature of this object, we do not include it in our analysis here, though we show the {\it HST} image and SN location in Appendix~\ref{sec:10afx}. Additionally, the SLSN PS1-10ky \citep{ccs+11} was targeted as part of program GO-13022 but its host galaxy was not detected. It is therefore not included in the discussion, except for where the upper limit is relevant.

In addition to the {\it HST} data from our PS1 sample, a few SLSN host galaxies have {\it HST} images available in the public archive from proposal GO-13025 (PI: Levan). We included available public images covering rest-frame UV of 5 H-poor SLSNe in our analysis; targets and details of the observations are listed in Tables~\ref{tab:targets} and \ref{tab:hst}. As the public objects are generally at lower redshifts than the PS1 sample, most of these objects were imaged with the Wide Field Camera 3 (WFC3/UVIS). Unlike ACS, WFC3 images are not currently corrected for CTE losses as part of the standard {\it HST} pipeline, and we used the CTE correction software available from the WFC3 tools webpage\footnote{\url{http://www.stsci.edu/hst/wfc3/tools/cte\_tools}} before processing and combining the individual images with Astrodrizzle. Again, we used a {\tt pixfrac} of 0.8 and a final {\tt pixscale} of half the native image scale, which corresponds to a final scale of 0.0198\arcsec/pixel. All final drizzled images are shown in Figure~\ref{fig:hstpix}.

\subsection{Astrometry}
\label{sec:astrometry}

\subsubsection{PS1-HST Astrometry}
To determine the locations of the SNe relative to the {\it HST} high-resolution images, we take advantage of the astrometry framework in the PS1/MDS {\tt photpipe} pipeline. We use {\tt SExtractor}\footnote{\url{http://sextractor.sourceforge.net}} to create a catalog of suitable astrometric reference sources from the {\it HST} images, and shift a PS1 template image to this reference frame. Typically, there are 20-60 tie objects available between the ACS images and the PS1 templates, giving resulting tie uncertainties of 10-30~mas. These uncertainties are quoted as $\sigma_{\rm tie}$ in Table~\ref{tab:results}.

In addition to the uncertainty from the astrometric tie between the {\it HST} images and PS1 templates, there is a contribution based on how well we can determine the SN centroids. To calculate this, we re-run the supernova images through the pipeline, updating the astrometry of the nightly and reference images to the {\it HST}-defined reference frame, performing image subtractions, and calculate the SN centroids and associated error from the subtracted images. \citet{rsf+14} describe in detail how the astrometry is performed by the pipeline. We then combine all the individual measurements and calculate the weighted, 3$\sigma$-clipped average centroid for each SN. We find typical uncertainties in the SN centroids of 10-30~mas, depending on both the number of images available for each SN, and the seeing and signal-to-noise ratio (S/N) for the individual detections. We note that the scatter in the position measurements are comparable to the astrometric errors calculated by the pipeline (i.e., reduced $\chi^2$-values for the mean centroid close to 1). The uncertainties in SN positions are listed as $\sigma_{\rm SN}$ in Table~\ref{tab:results}. The total uncertainty of the SN position in the {\it HST} image is found by combining $\sigma_{\rm tie}$ and $\sigma_{\rm SN}$ in quadrature, and is depicted as red circles in Figure~\ref{fig:hstpix}.

\subsubsection{HST-HST Astrometry}
Two objects, PS1-11aib and SCP06F6, have {\it HST} images available for both the SLSNe and their host galaxies: PS1-11aib had both SN and host galaxy imaging done as part of program GO-12529, and SCP06F6 was discovered as part of the \textit{HST} Cluster Supernova Survey (program GO-10496; \citealt{bdt+09}).  In these two cases, we align the {\it HST} images of the SNe directly to the galaxy images, using the Drizzlepac task {\tt tweakreg}. Since there are not many stars in these fields, we again use catalogs of suitable sources created with {\tt SExtractor} as input (rather than the built-in {\tt imagefind} routine in Drizzlepac). We calculate the weighted SN centroids in the frame of the host galaxy; the $\sigma_{\rm SN}$ values quoted in Table~\ref{tab:results} is the total final position uncertainty.

\subsubsection{Literature Objects}
For the remaining literature objects, we use available ground-based imaging of the SNe for astrometry: PS1 3Pi images of SN\,2011ke and SN\,2012il \citep{isj+13}, GMOS imaging of SN\,2010gx \citep{psb+10}, and Liverpool Telescope images of SN\,2007bi \citep{gmo+09, ysv+10}. Here, we use {\tt SExtractor} to create catalogs of overlapping sources, and the IRAF task {\tt ccmap} to compute the astrometric tie for each SN image. As with the PS1/MDS images, we then combine the measurements from individual SN images to calculate the weighted SN centroid in the frame of the {\it HST} images. 

Unlike for the PS1-HST astrometry, we do not use host-subtracted images to determine the SN centroid for these literature objects, as we do not have host galaxy images in the same filters available. This could in principle bias the position measured. To minimize such effects, we use SN images as close to maximum light as possible, where the SNe are $\gtrsim 2-4~{\rm mag}$ brighter than their host galaxies, and the contribution of host galaxy flux to the measured centroid should therefore be negligible.

\section{Comparison Samples}
\label{sec:stats}
\begin{figure}
\begin{center}
\includegraphics[width=3.5in]{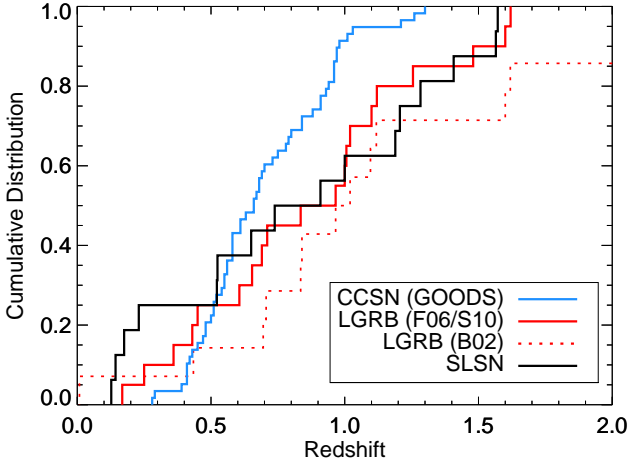}
\caption{Redshift distributions of our SLSN sample (black), and of the main comparison samples. The blue and red solid lines show the GOODS CCSN sample and LGRB sample presented in \citet{fls+06} and \citet{slt+10}. The red dotted line shows the redshift distribution of the LGRB sample from \citet{bkd02}, which is used in the offset comparison.
\label{fig:zdist}}
\end{center}
\end{figure}

We compare the locations of SLSNe within their host galaxies to different types of astrophysical transients. As our SLSN sample spans a wide range of redshifts ($0.1 \lesssim  z \lesssim 1.6$), the comparison samples should ideally cover a similar redshift range (to minimize effects due to galaxy evolution), and also come from an untargeted survey (to avoid biasing toward specific galaxy types). We use the GOODS sample of core-collapse SNe \citep{fls+06,slt+10} as our main SN comparison sample, as it satisfies both these criteria (see Figure~\ref{fig:zdist} for a redshift distribution comparison). Moreover, the SN locations within the host galaxies are very well determined as both the SNe and the host galaxies were observed with {\it HST}. One drawback of the GOODS sample is that most of the SNe in the core-collapse sample were not spectroscopically confirmed. As the main goal of the GOODS SN search was to find Type Ia SNe, the CCSN sample consists of SNe with colors incompatible with being Type Ia SNe (and therefore not followed up further), or SNe with spectra that were not Type Ia \citep{srd+04}. There is therefore no breakdown of sub-types within the GOODS sample. In addition, the papers describing the GOODS sample do not discuss host-SN offsets. 

For studies of spectroscopically confirmed CCSNe, as well as studies of host-SN offsets, then, we are limited to low-redshift samples. We utilize the studies of \citet{psb08}, \citet{kkp08} and \citet{kk12} for host-SN offsets and light distribution statistics for different types of SNe. In addition to the redshift difference to the SLSNe, these comparisons are complicated by the fact that the low-redshift samples contain a larger fraction of high-luminosity galaxies because the SNe in the samples come from targeted surveys.

LGRBs offer another interesting comparison sample to H-poor SLSNe. The two types of transients share a number of properties; both are rare and energetic explosions, with rates $\lesssim 10^{-3}$ of the CCSN rate \citep{qya+13, msr+14,wp10}. Like H-poor SLSNe, the SNe that accompany LGRBs are stripped of hydrogen. Moreover, \citet{lcb+14} found that their environments are similar on a galaxy-scale level, with both samples preferentially found in dwarf galaxy environments with low metallicities and high sSFRs. Locations of LGRBs within their host galaxies were studied in \citet{fls+06} and \citet{slt+10}; we will also refer to \citet{bkd02} for offsets of LGRBs. Figure~\ref{fig:zdist} also shows the redshift distributions of both these LGRB samples, which are well-matched to the SLSN sample.

To compare the distribution of SLSN properties to the other samples, we use the Kolmogorov-Smirnov (K-S) test. This test calculates the KS statistic $D$, defined as $D = \sup_x | F_1(x) - F_2(x)|$, where $F_1$ and $F_2$ are the empirical cumulative distribution functions of the two samples. This statistic is then compared to a theoretical distribution to calculate the probability that the two samples were drawn from the same underlying distribution.

\section{Galaxy Properties from Resolved Imaging}
\label{sec:galscale}
Before we turn to an analysis of the SN locations, discussed in Section~\ref{sec:results}, we investigate the morphologies of the SLSN host galaxies. This is helpful both for understanding the nature of the galaxies, and to normalize the SLSN offset measurements. 

\subsection{Galaxy Morphologies}
 Visual inspection of Figure~\ref{fig:hstpix} reveals that most SLSN host galaxies have irregular morphologies. There is a striking lack of grand design spiral galaxies compared to the host galaxies of normal SNe: the GOODS survey found that approximately half of the CCSNe in this redshift range are found in massive spiral galaxies \citep{fls+06,slt+10}. In contrast, none of the SLSN hosts in our sample have visible spiral structure. In addition, the deficit of spiral galaxy hosts agrees with the results from \citet{lcb+14} that SLSNe select different environments from normal CCSNe, suggesting that additional factors beyond star formation is necessary to produce a SLSN. Moreover, this is another characteristic that SLSN host galaxies share with LGRB host galaxies, which are also deficient in bright spirals and generally found in galaxies with irregular morphologies \citep{fls+06, wbp07}.

Another interesting feature of Figure~\ref{fig:hstpix} is that roughly half of the galaxies exhibit a morphology that is either asymmetric, off-center or consisting of multiple peaks. Such morphologies are common also amongst LGRB host galaxies: \citet{wbp07} classified $\gtrsim 60\%$ of the galaxies in their sample as either showing features consistent with merging system, or asymmetric and irregular structure.

\subsection{Galaxy Sizes}
\label{sec:size}
We use {\tt SExtractor} to measure effective galaxy radii, using a S/N $> 1$ criterion to determine which pixels are part of the galaxy. In Table~\ref{tab:results} we list $r_{50}$ and $r_{80}$, the radii estimated by {\tt SExtractor} to contain 50 and 80\% of the total host light, respectively. PS1-11aib is unresolved in the {\it HST} image, and we take the FWHM of the point-spread function (PSF) as an upper limit on its size.

\begin{figure}
\begin{center}
\includegraphics[width=3.5in]{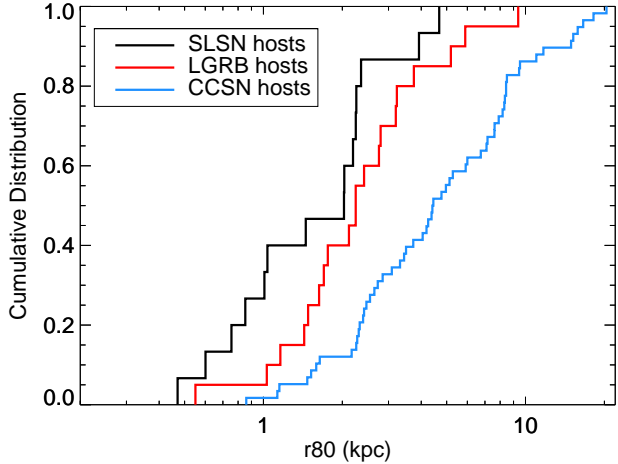}
\caption{Distribution of $r_{80}$, the radius containing 80\% of the galaxy flux, for the SLSN host sample (black), LGRB host galaxies (red) and core-collapse SN host galaxies from GOODS (blue).
\label{fig:r80}}
\end{center}
\end{figure}

As is also evident from the images in Figure~\ref{fig:hstpix}, the SLSN host galaxies are remarkably compact. The half-light radii span 0.2 to 2.9~kpc, with a median of 0.9~kpc. Figure~\ref{fig:r80} shows the distribution of $r_{80}$, the radius containing 80\% of the total light, compared to the LGRB and GOODS CCSN host samples from \citet{slt+10}. The SLSN host sizes are comparable to the LGRB hosts, if overall slightly smaller, and their distributions are statistically compatible. In contrast, the GOODS CCSN host galaxies are significantly larger, with a median $r_{80}$ of 4.45~kpc. A KS test rejects the null hypothesis that the two samples are drawn from the same distribution at high significance ($p = 2 \times 10^{-5}$).

Since our observations probe the rest-frame UV, which traces star formation, one might worry that the small sizes measured are a result of only detecting bright knots of star formation rather than the overall distribution of stars; for example in the image of SN\,2012il (Figure~\ref{fig:hstpix}), an extended structure at lower surface brightness is visible, and may be more representative of the true size of the galaxy than the two bright knots in the left of the image that dominate the UV light. One way to test this is to compare images that trace rest-frame optical or IR light to our UV images. Such images are available for about half the sample. For SN\,2012il, the IR extent of the host galaxy is indeed similar to the low surface brightness component seen in the UV; the host of SN\,2007bi also appears to be more extended in the IR though at a low S/N level. The remaining five galaxies with WFC3/IR imaging in our sample show similar morphologies and sizes as in the UV, suggesting that the UV images are generally representative of the overall host galaxy size. We also note that based on spectral energy distribution (SED) modeling \citep{lcb+14}, we do not generally expect a significant component of old stars.

\subsection{Star Formation Rates and SFR Surface Density}
Since our images cover rest-frame UV, we can use the galaxy fluxes to estimate UV-derived SFRs. We use the ``isocorr'' magnitudes returned by {\tt SExtractor} as the estimate of the total flux from the galaxy, and convert the UV luminosity into SFRs using the relation from \citet{ken98}: ${\rm SFR}~(M_{\odot}/{\rm yr}) = 1.4 \times 10^{-28} L_{\nu, {\rm UV}} ({\rm erg}~{\rm s}^{-1}{\rm Hz}^{-1})$. 

\begin{figure}
\begin{center}
\includegraphics[width=3.5in]{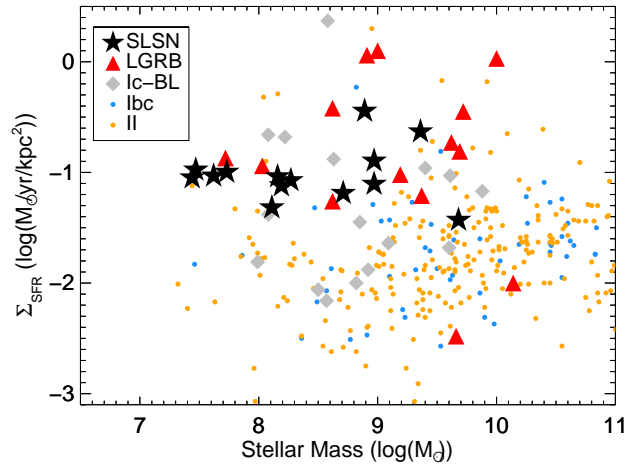}
\caption{SFR surface density as a function of stellar mass. The SLSN host galaxies are shown as black stars; data for other types of transients are taken from \citet{kfm+14}. Similar to the host galaxies of Type Ic-BL SNe and LGRBs, the SLSN host galaxies have high SFR surface densities given their stellar masses.
\label{fig:sfr_sigma}}
\end{center}
\end{figure}

Using the galaxy sizes we also calculate the SFR surface density, i.e. the SFR per unit area. We use the isophotal area determined by {\tt SExtractor} as our best estimate of the total area of the galaxies.  In Figure~\ref{fig:sfr_sigma} we plot the SFR density as a function of stellar mass, using the masses derived from SED fitting by \citet{lcb+14}, except for SN\,2007bi where we use the updated mass from \citet{csj+14}. Data for the host galaxies of other types of transients are taken from \citet{kfm+14}. We find that, like LGRBs and Type Ic -BL (broad-lined) SNe, the SLSN host galaxies have high SFR densities for their stellar mass, ranging from 0.04 to 0.4~$M_{\odot}{\rm~ yr}^{-1}{\rm~kpc}^{-2}$ with a median value of 0.09~$M_{\odot}{\rm~ yr}^{-1}{\rm~kpc}^{-2}$.

For the undetected galaxy in our sample, PS1-10ky, we measure the standard deviation of the background at the position of the transient, and use this to calculate an upper limit on the galaxy surface brightness. At the redshift of PS1-10ky, the resulting limit on SFR density is  $ 0.1 M_{\odot}{\rm~ yr}^{-1}{\rm~kpc}^{-2}$ ($3\sigma$), similar to the lowest-mass detections. Note that this limit assumes the galaxy is at least as large as the PSF of the image ($0.075$\arcsec, corresponding to $\sim 600~{\rm pc}$); if the galaxy is smaller it could in principle have a higher SFR density than this limit and remain undetected.

We caution that the samples we are comparing to used sizes derived from rest-frame optical rather than UV images, which could lead to higher derived SFR densities if the UV emission is not representative of the true size of the galaxy. To follow the same relation that is seen in normal core-collapse SN host galaxies, however, the SLSN sizes would have to be underestimated by a factor of $\sim 10$, which is not supported by the subsample for which we have rest-frame optical or IR imaging (Section~\ref{sec:size}).

\section{Supernova Locations}
\label{sec:results}
\subsection{Offsets}
\label{sec:offsets}
To calculate the offset from the SLSN locations to the host galaxy, we first need to define the center of each galaxy. We use the centers output by {\tt SExtractor}, which correspond to the flux-weighted galaxy centroids. Given the irregular morphology of many of our targets, we note that the center calculated in this way does not necessarily correspond to the brightest region of the galaxy. The uncertainty in the host galaxy centroid is listed as $\sigma_{\rm gal}$ in Table~\ref{tab:results}.

\begin{figure*}
\begin{center}
\begin{tabular}{cc}
\includegraphics[width=3.5in]{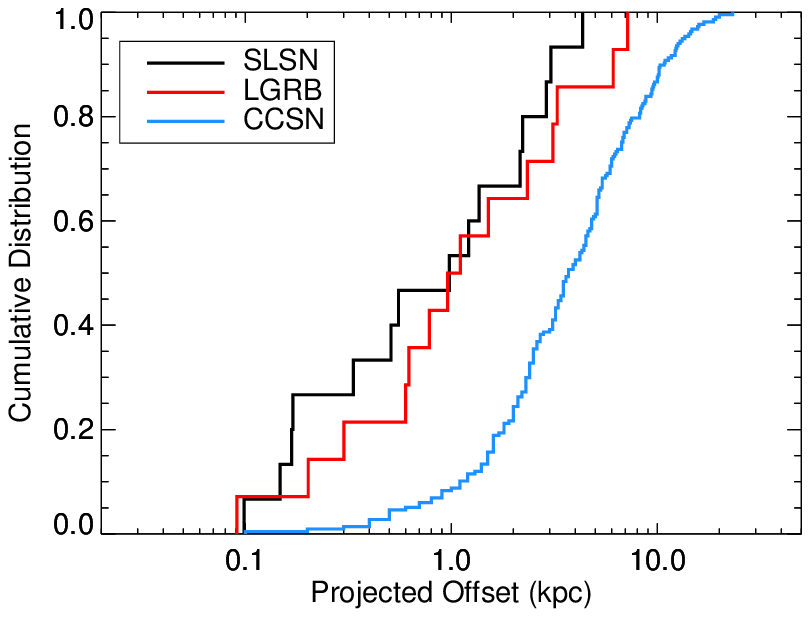} & \includegraphics[width=3.5in]{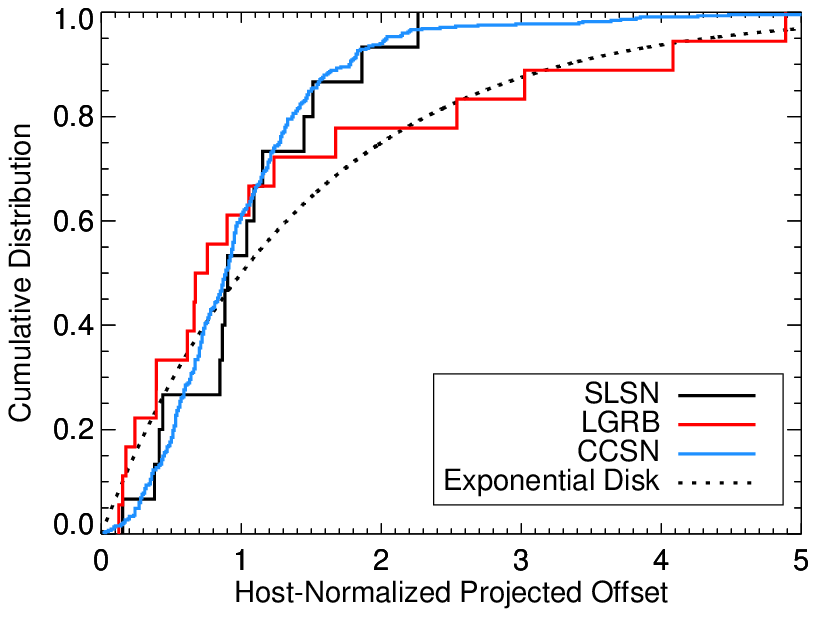}
\end{tabular}
\caption{Distributions of projected offsets from the host center, both in physical units (left) and normalized by the host galaxy's half-light radius (right). The SLSN sample is shown in black, LGRBs from \citet{bkd02} in red, and low-z core-collapse SNe from \citet{psb08} and \citet{kk12} in blue. Also shown is the expected distribution for an exponential disk model (dashed line).
\label{fig:offsets} }
\end{center}
\end{figure*}

The distribution of resulting offsets is shown in Figure~\ref{fig:offsets}, both in projected kpc (left) and normalized by the host galaxy size (right). We here use $r_{50}$ rather than $r_{80}$, since this is what was used to normalize the offsets in the comparison samples. The SLSNe have offsets ranging from 0.1 to 4.3~kpc, with a median of 1.0~kpc, comparable to the LGRBs and significantly smaller than the offsets of the low-redshift CCSNe. However, when normalized by host galaxy size, the SLSNe are statistically compatible to the two other populations, and in particular all three populations have median offsets of about one half-light radius. This indicates that the SLSN locations overall track the radial distribution of UV light, similar to other transients with massive star progenitors.

\subsection{Light Distribution Analysis}
\label{sec:light_dist}
The offset technique is limited in comparing the locations of transients to the overall light distribution because many of the SLSN host galaxies exhibit irregular, asymmetric structure. The distance from the flux-weighted center of the galaxy is therefore not necessarily a good indicator of the flux level at the SN position. For this reason, \citet{fls+06} developed a morphology-independent technique for quantifying the extent to which SNe trace their host light distribution, by determining the pixel on which the SN occurred, and computing the fraction of light in galaxy pixels of lower surface brightness. A hypothetical population of sources that perfectly tracked the underlying light distribution would follow a uniform distribution: a pixel with twice as much flux would, statistically speaking, be exactly twice as likely to contain a transient.

To calculate this statistic, we first determine which pixels are part of the galaxy. We fit a Gaussian profile to the sky brightness distribution near the galaxy, and determine the 1$\sigma$ cutoff level (equivalent to taking S/N $> 1$). Consecutive pixels above this cutoff level are then defined to be part of the galaxy, and we compute the fraction of light in galaxy pixels fainter than the SLSN position. In cases where the SN position is known to a precision worse than the FWHM of the image, we first convolve the image with a Gaussian of the same width as the SN position uncertainty. The results for each galaxy are listed in Table~\ref{tab:results}.

\begin{figure}
\begin{center}
\includegraphics[width=3.5in]{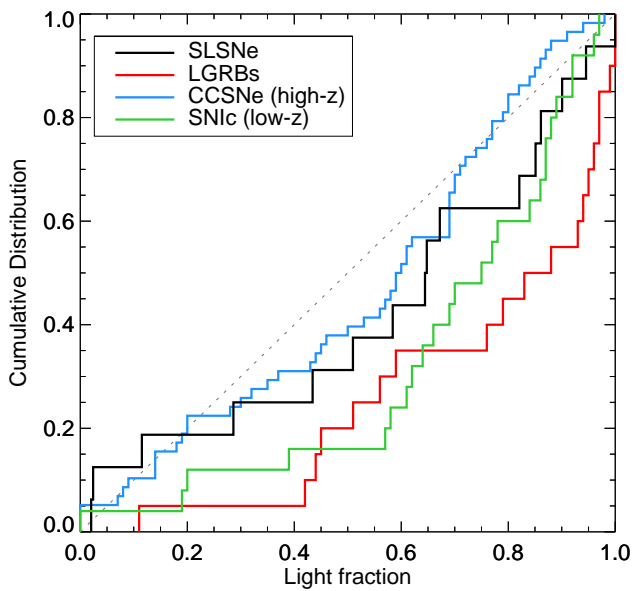}
\caption{Cumulative distribution of the fraction of total light found in pixels fainter than the location of the transient. The SLSN sample is shown in black, LGRBs in red, GOODS CCSNe in blue and local Type Ic SNe in green. The dotted line marks a uniform distribution, expected for a theoretical population that perfectly tracked its host light.
\label{fig:light_dist} }
\end{center}
\end{figure}

In Figure~\ref{fig:light_dist} we show the cumulative distribution of the light fractions of SLSN locations, compared to the GOODS CCSNe, LGRBs \citep{slt+10} and local Type Ic SNe \citep{kkp08}. The dashed line marks the expectation of a uniform distribution. The SLSN locations overall are slightly skewed toward brighter pixels, with a median value is 0.65. This suggests that SLSNe are indeed correlated with the UV light of their host galaxies, and are slightly more likely to be found in brighter regions of their host galaxies. They do not appear to be as strongly correlated with the brightest regions as are LGRBs, which have a median value of 0.88. In particular, only one ($6 \%$) of the LGRBs in the sample from \citet{slt+10} are found in regions of fractional brighness $< 40 \%$, whereas about a quarter of the SLSNe are found at such low flux values. The sample sizes are small, however, and statistically we cannot rule out either that the SLSNe come from the same distribution as the LGRBs ($p = 0.25$), or that they are drawn from a uniform distribution ($p = 0.44$).

The comparison is complicated by the fact that the SLSN host galaxies are overall fainter than either of the galaxy populations we are comparing to, and so it is possible the SLSN distribution is shifted to lower relative flux values due to the lower-surface brightness parts of the galaxies not being above the noise threshold. This effect is not likely to be severe, however, since the faintest pixels in the distribution contribute little to the overall {\it total} flux, which is what matters the most for the relative position. This effect was considered in detail by \citet{fls+06} when comparing the LGRB sample to the GOODS CCSNe, by experimenting with artificially increasing the noise in their images by a magnitude (thus losing a larger fraction of the galaxy edge flux). They found that their results were overall unaffected. It is also worth noting that the galaxies where we find the lowest fractional flux levels at the SN positions are in fact some of the brightest galaxies in the sample: PS1-12bqf, PS1-11bam and PS1-12bmy (Figure~\ref{fig:hstpix}; Table~\ref{tab:results}). Therefore, the result that some SLSNe explode in regions of their galaxy with very little UV flux is unlikely to be caused by a surface brightness bias.

Another potential bias arises due to that the contrast between a typical SLSN and its host galaxy is several magnitudes greater than that between a typical CCSN and its host. As a result, the SLSNe are relatively easier to detect in the brightest regions of their host galaxies, whereas the CCSN sample may be missing events in the bright galaxy cores. \citet{fls+06} examined this effect in the GOODS sample, and estimated that the fraction of central SNe missed in the GOODS sample was $< 10\%$. In addition, the low-z Type Ic SN sample \citep{kkp08} appears to trace the brightest regions of their hosts at least as strongly as the SLSNe do. Thus, SN-host contrast is unlikely to be affecting the comparison on a significant level.

\section{Discussion}
\label{sec:disc}
Both the offset distribution and the fractional flux distribution of SLSNe suggest that their locations are correlated with the UV light and hence with recent star formation activity. This is further evidence that SLSNe come from massive stars, as is also seen by their associations with star-forming galaxies in general, and high sSFRs in particular \citep{lcb+14}.  While their locations are statistically consistent with those of LGRBs, the SLSNe appear to not be as strongly correlated with the brightest regions of their host galaxies as the LGRBs are (Figure~\ref{fig:light_dist}). Since a stronger correlation with star formation tracers is generally interpreted as evidence for a younger and more massive progenitor population \citep{fls+06,ahj+12,kkp08}, the simplest interpretation is that SLSN progenitors are older/less massive stars than LGRB progenitors. We note that recently, \citet{lsk+14} have argued that SLSNe result from more massive stars than LGRBs; our result of SLSN locations is potentially at odds with this interpretation. Instead, as we argued in \citet{lcb+14}, the host galaxy properties seem to lead to the conclusion that the progenitors require low metallicity, potentially as a requirement for a magnetar remnant.

While the locations of SLSNe may not be more strongly correlated with star formation activity than those of ordinary core-collapse SNe, the fact remains that their overall host environments are strikingly different. Our results can also be interpreted to mean that the most important environmental factor for producing a SLSN is something different than the progenitor mass alone. One such factor could be metallicity, as H-poor SLSNe are generally found in low-metallicity galaxies \citep{lcb+13, lcb+14, csb+13, csj+14, lsk+14}. In addition, several proposed models for SLSNe are consistent with a low metallicity preference, either directly in the case of PISN models, or indirectly in the case of models that require a rapidly rotating progenitor, such as the magnetar model \citep{cw12a,cw12b, woo10, kb10}. A low-metallicity requirement could explain why some SLSNe appear to explode in regions away from the most intense star formation, especially in the more massive galaxies in our sample. In particular it is worth noting that in all cases where the SLSN exploded in what looks like an ordinary disk galaxy (PS1-12bqf in the {\it HST} sample, and the lower-redshift PTF11rks and MLS121104 in \citealt{lcb+14}), the SN location is in the outskirts of the galaxy. The fact that our sample is dominated by irregular dwarf galaxies could explain why a potential metallicity preference is not reflected as a preference for large offsets in the overall distribution.

While metallicity is one possible option (and one that is often invoked also to explain the environmental preferences of LGRBs), it is not the only possibility. Recently, \citet{kfm+14} showed that the host galaxies of both broad-lined type Ic SNe and LGRBs have high stellar mass and star formation densities, compared to SDSS galaxies of similar masses, and we find that SLSN host galaxies show the same trend (Figure~\ref{fig:sfr_sigma}). Since this trend cannot be explained by a simple metallicity preference, \citet{kfm+14} instead argue the key factor may be that massive binary progenitor systems can form more efficiently in regions of dense star formation \citep{gbk10,sab13}. \citet{vp13} have proposed a dynamical origin for SLSNe, as the result of runaway collisions in dense star clusters. If this is the dominant channel for producing SLSNe, however, it would be difficult to explain our result that some fraction of SLSNe explode well away from the brightest regions of star formation.

In the context of the interaction-driven model for SLSNe, the natural comparison sample would be the H-rich SLSNe, since these do show clear signs of interaction in their spectra. No similar study of the locations of this class has yet been carried out, unfortunately, though their overall environments appear to be less extreme than those of H-poor SLSNe \citep{lsk+14}. For interaction-driven SNe in general, \citet{haj+14} examined the locations of 26 Type IIn SNe, and found that they overall traced the UV light distribution, but was not correlated with ongoing star formation as traced by H$\alpha$. This result is perhaps surprising, as Type IIn SNe are thought to have massive progenitors such as Luminous Blue Variable stars (LBVs; e.g. \citealt{gl09}), whereas the Type IIn locations would suggest similar progenitor masses as ordinary Type IIp SNe. Recently, \citet{st15} have argued that both the observed isolated locations of LBVs and the locations of Type IIn SNe are naturally explained in a scenario where LBVs are the product of binary evolution, with a significant fraction being kicked from its birth cluster when the companion goes supernova. This illustrates how other factors than progenitor mass can have a significant impact on the observed locations of SN subtypes, and that the different correlations between SN locations and UV light between different populations need not be due to a simple mass difference if binary progenitor models are considered.

\section{Conclusions}
\label{sec:conc}
We have carried out the first study of the sub-galactic environments of H-poor SLSNe, using resolved rest-frame UV imaging from {\it HST} and precise SN locations. Our conclusions are as follows:
 \begin{itemize}

\item The host galaxies of H-poor SLSNe are generally irregular dwarf galaxies:  about half show asymmetric morphology or multiple peaks, and there is a distinct lack of grand-design spiral galaxies compared to CCSN host galaxies in the same redshift range.

\item SLSN host galaxies are compact, with a median half-light radius of 0.9~kpc. The median SFR surface density, as derived from the UV flux, is 0.1~$M_{\odot}{\rm yr}^{-1}{\rm kpc}^{-2}$. SLSN host galaxies exhibit high SFR surface densities for their stellar masses compared to host galaxies of local CCSNe. 

\item The median offset between SLSNe and their hosts is $\sim$ 1~kpc. The normalized offset distribution is consistent with those of other types of transients with massive star progenitors, with a median normalized offset of one half-light radius.

\item The locations of H-poor SLSNe are correlated with the UV light. We find that their distribution is intermediate between those of LGRBs (which trace the brightest regions of their hosts) and a uniform distribution, and cannot be distinguished statistically from either with the current sample size.

\end{itemize}

The galaxy properties derived from the {\it HST} images support the overall picture that H-poor SLSNe explode in host galaxies that are overall different from core-collapse SN hosts, and that share many similarities with LGRB host galaxies. Both samples are primarily found in irregular galaxies, with similar typical sizes. In addition, both are found in galaxies with high star formation surface densities. Thus, the galaxy-scale properties support the results of \citet{lcb+14} that similar environmental factors are necessary for stars to end their lives as either a H-poor SLSN or a LGRB. 

At the same time, our study of the sub-galactic locations of SLSNe indicates that the {\it local} environments of SLSNe and LGRBs may be different, with SLSNe being less biased toward the brightest regions of their host galaxies (although a larger sample size is necessary to distinguish them statistically). This can be interpreted as SLSN progenitors being less massive and longer-lived stars than LGRB progenitors, contrary to recent claims that SLSNe are the very first stars to explode in a starburst \citep{lsk+14}. Our results are consistent with the recent results of \citet{vsg+14}, however, who find that the ISM column densities along SLSN lines of sight are on the low end of what is seen in LGRBs, also suggesting that they trace different local environments.

\acknowledgements
We thank the anonymous referee for helpful comments that improved the clarity of the paper. Support for programs number GO-12529, GO-13022 and GO-13326 was provided by NASA through a grant from the Space Telescope Science Institute, which is operated by the Association of Universities for Research in Astronomy, Inc., under NASA contract NAS5-26555. The Pan-STARRS1 Surveys (PS1) have been made possible through contributions of the Institute for Astronomy, the University of Hawaii, the Pan-STARRS Project Office, the Max-Planck Society and its participating institutes, the Max Planck Institute for Astronomy, Heidelberg and the Max Planck Institute for Extraterrestrial Physics, Garching, The Johns Hopkins University, Durham University, the University of Edinburgh, Queen's University Belfast, the Harvard-Smithsonian Center for Astrophysics, the Las Cumbres Observatory Global Telescope Network Incorporated, the National Central University of Taiwan, the Space Telescope Science Institute, the National Aeronautics and Space Administration under Grant No. NNX08AR22G issued through the Planetary Science Division of the NASA Science Mission Directorate, the National Science Foundation under Grant No. AST-1238877, the University of Maryland, and Eotvos Lorand University (ELTE) and the Los Alamos National Laboratory. This paper includes data based on observations made with the NASA/ESA {\it Hubble Space Telescope} and obtained from the Hubble Legacy Archive, which is a collaboration between the Space Telescope Science Institute (STScI/NASA), the Space Telescope European Coordinating Facility (ST-ECF/ESA) and the Canadian Astronomy Data Centre (CADC/NRC/CSA). SJS acknowledges funding from the European Research Council under the European Union's Seventh Framework Programme (FP7/2007-2013)/ERC grant agreement n$^{\rm o}$~[291222]. RPK acknowledges support from the National Science foundation through AST1211196.


\begin{deluxetable}{lccc}
\tablecaption{Target List}
\tablehead{
\colhead{Object} &
\colhead{Redshift} &
\colhead{RA} &
\colhead{Dec} 
}
\startdata
PS1-12bqf &  $0.522$   & \ra{02}{24}{54.621}  & \dec{-04}{50}{22.72} \\
PS1-11ap  &  $0.524$   & \ra{10}{48}{27.752}  & \dec{+57}{09}{09.32} \\
PS1-10bzj  &  $0.650$   & \ra{03}{31}{39.826}  & \dec{-27}{47}{42.17} \\
PS1-11bdn &  $0.738$   & \ra{02}{25}{46.292}  & \dec{-05}{03}{56.57} \\
PS1-10awh  &  $0.909$   & \ra{22}{14}{29.831}  & \dec{-00}{04}{03.62} \\
PS1-10ky   &  $0.956$   & \ra{22}{13}{37.851}  & \dec{+01}{14}{23.57} \\
PS1-11aib  &  $0.997$    & \ra{22}{18}{12.217}  & \dec{+01}{33}{32.01} \\
PS1-10pm   &  $1.206$   & \ra{12}{12}{42.200}  & \dec{+46}{59}{29.48} \\
PS1-11tt   &  $1.283$   & \ra{16}{12}{45.778}  & \dec{+54}{04}{16.96} \\
PS1-10afx &  $1.388$   & \ra{22}{11}{24.160}  & \dec{+00}{09}{43.49} \\
PS1-11afv  &  $1.407$   & \ra{12}{15}{37.770}  & \dec{+48}{10}{48.62} \\
PS1-11bam &  $1.565$   & \ra{08}{41}{14.192}  & \dec{+44}{01}{56.95} \\
PS1-12bmy &  $1.572$   & \ra{03}{34}{13.123}  & \dec{-26}{31}{17.21} \\
SN\,2007bi & $0.128$  & \ra{13}{19}{20.19}    & \dec{+08}{55}{44.3} \\
SN\,2011ke & $0.143$  & \ra{13}{50}{57.78}   & \dec{+26}{16}{42.40} \\
SN\,2012il & $0.175$  & \ra{09}{46}{12.91} & \dec{+19}{50}{28.7} \\
SN\,2010gx & $0.230$   & \ra{11}{25}{46.71} & \dec{-08}{49}{41.4} \\ 
SCP06F6    & $1.189$   & \ra{14}{32}{27.395}  & \dec{+33}{32}{24.83} 
\enddata
\label{tab:targets}
\end{deluxetable}

\begin{deluxetable}{lccccc}
\tablecaption{Summary of HST Observations}
\tablehead{
\colhead{Object} &
\colhead{UT Date} &
\colhead{Instrument} &
\colhead{Filter} &
\colhead{Rest-frame $\lambda_{\rm eff}$} &
\colhead{Exposure Time} \\
\colhead{} &
\colhead{(YYYY-MM-DD)} &
\colhead{} &
\colhead{} &
\colhead{(\AA)} &
\colhead{(s)}
}
\startdata
PS1-12bqf & 2013-11-18 &  ACS/WFC & F475W & 3118    & 2200 \\
PS1-11ap  & 2013-10-09 &  ACS/WFC & F475W & 3113    & 2464 \\
PS1-10bzj\tablenotemark{a}  & 2002-11-11 &  ACS/WFC & F606W & 3589  & 2160 \\
PS1-11bdn & 2013-11-13  &  ACS/WFC & F475W & 2730  & 2200 \\
PS1-10awh & 2013-09-04  &  ACS/WFC & F606W & 3102  & 680  \\
PS1-10ky  & 2012-12-13  &  ACS/WFC & F606W & 3027  & 680  \\
PS1-11aib & 2013-09-12  &  ACS/WFC & F625W & 3160  & 1000 \\
PS1-10pm  & 2012-12-10  &  ACS/WFC & F606W & 2684  & 1960 \\
PS1-11tt  & 2012-12-02  &  ACS/WFC & F606W & 2593  & 1960 \\
PS1-10afx & 2013-10-08  &  ACS/WFC & F814W & 3373  & 2200 \\
PS1-11afv & 2013-04-09  &  ACS/WFC & F606W & 2460  & 1960 \\
PS1-11bam & 2013-10-11  &  ACS/WFC & F814W & 3141  & 2304 \\
PS1-12bmy & 2013-09-17  &  ACS/WFC & F814W & 3132  & 2224 \\
SN\,2007bi & 2012-11-27 & WFC3/UVIS & F336W & 2974    & 1808 \\
SN\,2011ke & 2013-05-16 & WFC3/UVIS & F336W & 2935    & 2044 \\
SN\,2012il & 2013-01-02 & WFC3/UVIS & F336W & 2855    & 2036 \\
SN\,2010gx & 2012-11-22 & WFC3/UVIS & F390W &  3190   & 1808 \\
SCP06F6   & 2013-05-23  &  ACS/WFC  & F606W &  2705   & 8054
\enddata
\tablenotetext{a}{Archival data from the GEMS survey \citep{rbb+04}.}
\label{tab:hst}
\end{deluxetable}

\begin{deluxetable}{lccccccccc}
\tablecaption{Results}
\tablehead{
\colhead{Object} &
\colhead{$\sigma_{\rm tie}$} &
\colhead{$\sigma_{\rm SN}$} &
\colhead{$\sigma_{\rm gal}$} &
\colhead{$r_{50}$} &
\colhead{$r_{80}$} &
\colhead{$\log(\Sigma_{\rm SFR})$} &
\colhead{Projected Offset} &
\colhead{Normalized Offset} &
\colhead{Light Fraction} \\
\colhead{} &
\colhead{(mas)} &
\colhead{(mas)} &
\colhead{(mas)} &
\colhead{(kpc)} &
\colhead{(kpc)} &
\colhead{($M_{\odot}{\rm yr}^{-1}{\rm kpc}^{-2}$)} &
\colhead{(kpc)} &
\colhead{($r/r_{50}$)} &
\colhead{} 
}
\startdata
PS1-12bqf  & 13.1 & 16.1    & 4.3 & 2.87 & 4.69 &   -1.4  & 4.34  & 1.5   & 0.12 \\
PS1-11ap   & 23.5 & 4.9     & 2.1 & 0.85 & 1.45 &   -1.2  & 0.98  & 1.2   & 0.65 \\
PS1-10bzj  & 31.4 & 33.9    & 1.5 & 0.58 & 1.01 &   -1.0  & 0.51  & 0.9   & 0.51 \\
PS1-11bdn  & 14.6 & 18.7    & 5.9 & 1.25 & 2.26 &   -1.3  & 1.36  & 1.1   & 0.58 \\
PS1-10awh  & 9.8  & 16.1    & 9.7 & 0.88 & 2.03 &   -1.0  & 0.33  & 0.4   & 0.43 \\
PS1-11aib  & \nodata & 12.3 & \nodata  & $ < 0.4$ & \nodata & \nodata & \nodata  & \nodata  & 1.00 \\
PS1-10pm   & 8.1  & 19.9    & 7.7 & 2.57 & 3.92  &  -1.1   &  2.22 &  0.9  & 0.29 \\
PS1-11tt   & 13.6 & 16.3    & 5.3 & 1.34 & 2.04  &  -1.0   & 1.21  &  0.9  & 0.90 \\
PS1-11afv  & 11.8 & 34.8    & 4.7 & 1.49 & 2.25  &  -0.9   & 2.15  &  1.4  & 0.64 \\
PS1-11bam  &  8.2 & 21.8    & 2.8 & 1.34 & 2.36  &  -0.4   & 3.04  & 2.3   & 0.02 \\
PS1-12bmy  & 10.5 & 36.8    & 4.1 & 1.55 & 2.20  &  -0.6   & 2.89  & 1.9   & 0.02 \\
SN\,2007bi & \nodata  &19.6 & 4.8 & 0.20 & 0.47  &   -1.0  &  0.17 &  0.8  & 0.67 \\
SN\,2011ke & 16.8 & 14.5    & 2.4 & 0.34 & 0.60  &   -1.0  & 0.15  & 0.4   & 0.86 \\
SN\,2012il & 36.5 & 26.0    & 3.3 & 0.53 & 0.85  &   -1.1  & 0.55  & 1.0   & 0.95 \\
SN\,2010gx & \nodata & 16.3 & 2.7 & 0.41 & 0.76  &  -1.1   & 0.17  & 0.4   & 0.82 \\
SCP06F6    & \nodata & 9.5  & 4.4 & 0.65 & 1.04  &  -1.4   & 0.10  & 0.2   & 0.85 
\enddata
\label{tab:results}
\tablecomments{$\sigma_{\rm tie}$ is the uncertainty in the astrometric tie between the {\it HST} image and a deep template, where applicable. $\sigma_{\rm SN}$ is the uncertainty in the SN centroid.  $\sigma_{\rm gal}$ is the uncertainty of the galaxy centroid.}
\end{deluxetable}

\appendix
\section{HST Image of the Host Galaxy of PS1-10afx}
\label{sec:10afx}

\begin{figure}
\begin{center}
\includegraphics{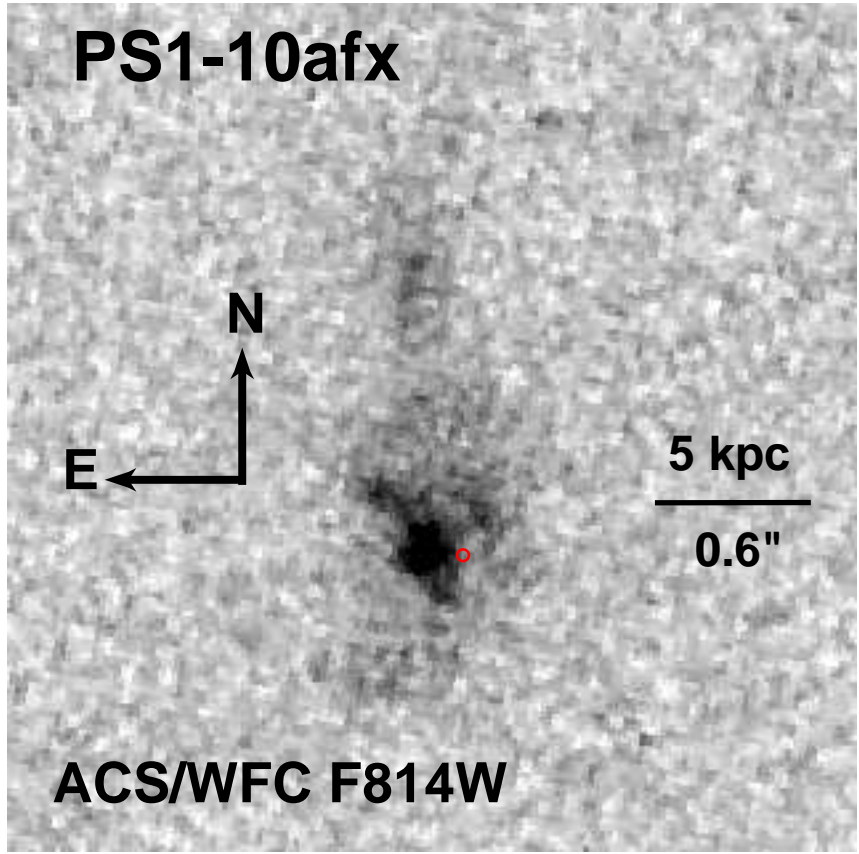}
\caption{$4\arcsec \times 4\arcsec$ image of the field around PS1-10afx. The SN location is marked with the red circle.
\label{fig:10afx}
}
\end{center}
\end{figure}

The unusual transient PS1-10afx was included in our {\it HST} sample because it was initially considered to be a SLSN at $z=1.388$ \citep{cbr+13}. However, \citet{qwo+13} identified it as a strongly gravitationally lensed SN Ia (magnification $\sim$30).  Subsequent spectroscopy by \citet{qom+14} identified an emission line from a fainter foreground galaxy at $z=1.1168$ along the line of sight to the brighter host galaxy. We present the F814W image of the field in Figure~\ref{fig:10afx} with the transient location marked by a red circle.  No arcs or other strongly distorted images of the background galaxy are clearly identifiable.  Deeper images with more color information would be necessary to decompose the observed source into a background host and foreground lens.  We leave an analysis of the implied constraints on the lensing geometry to future work.

\end{document}